\documentclass[preprint,12pt]{elsarticle}

\pdfoutput=1

\usepackage{graphics}
\usepackage{graphicx}
\usepackage{epsfig,verbatim}
\usepackage{dcolumn}
\usepackage{amssymb}
\usepackage{amsmath}
\usepackage[english]{babel}
\usepackage[latin1]{inputenc}
\usepackage{natbib}
\usepackage{epsfig}
\usepackage{bm}
\usepackage{eucal}
\usepackage{verbatim}
\usepackage{latexsym}
\usepackage{xcolor}
\usepackage{color}
\usepackage{subfigure}
\begin{document}

\begin{frontmatter}

\title{Reliability of the optimized perturbation theory in the 0-dimensional $O(N)$ scalar field model}

\author[ift]{D\'erick S. Rosa}
\ead{derick@ift.unesp.b}
\author[ufsm,kent]{R. L. S. Farias}
\ead{rfarias@kent.edu}
\author[uerj]{Rudnei~O.~Ramos}
\ead{rudnei@uerj.br}
\address[ift]{Instituto de F\'{\i}sica Te\'orica, Universidade Estadual
  Paulista, \\ Rua Dr. Bento Teobaldo Ferraz, 271 - Bloco II, 
  01140-070 S\~ao Paulo, SP, Brazil}
\address[ufsm]{Departamento de F\'isica, Universidade Federal de
  Santa Maria, \\
   97105-900 Santa Maria, RS, Brazil}
\address[kent]{Department of Physics, Kent State University, Kent, OH 44242, 
United States}   
\address[uerj]{Departamento de
  F\'{\i}sica Te\'orica, Universidade do Estado do Rio de Janeiro, \\ 20550-013
  Rio de Janeiro, RJ, Brazil}

\begin{abstract}
We address the reliability of the Optimized Perturbation Theory (OPT)
in the context of the 0-dimensional $O(N)$ scalar field model.  The effective
potential, the self-energy and the 1PI four-point Green's function
for the model are computed using different
optimization schemes and the results contrasted to the exact results
for the model. Our results are also compared to those obtained with
the $1/N$-expansion and  with  those from ordinary perturbation
theory. The OPT results are shown to be stable even at large couplings
and to have better convergence properties than the ones produced in
the $1/N$-expansion.  It is also shown that the principle of minimal
sensitive optimization procedure used in conjunction with the OPT
method  tends to always produce better results, in particular when
applied directly to the self-energy. 
\end{abstract}

\end{frontmatter}

\section{Introduction}

Perturbation theory is the most comprehensive way of studying
nonlinear problems in physics area wide. However, it is a fact that
not always we can rely on some small quantity in the theory which we
can use as a parameter which we can perturb physical quantities of
interest, or even when we do have, it is not warranted that a
perturbative series might be well posed, i.e., converge after a few
terms are considered. Most of the times we must make use of some
nonperturbative method to get around these problems.  One typical
example where perturbation theory breaks down  is in the studies of
phase transitions in general, particularly close to a critical
point. This can also happen due to the appearance of large
infrared divergences~\cite{Gross:1980br}, as in the case where
massless modes are present, or  close to a transition point in field
theories displaying a second order phase
transition~\cite{Espinosa:1992gq} or a weakly first order
transition~\cite{Gleiser:1992ed}.  In all these cases, the use of some
reliable nonperturbative technique is required to proper study these
systems. Among the analytical nonperturbative techniques, one can cite
for example making use of a discretization of the system and studying
it numerically (e.g., lattice simulations), make use of analytical
methods like an expansion in the number of field components, $N$, in
the case of field theory, using the
$1/N$-approximation~\cite{Moshe:2003xn}, among other methods.

In this work we want to access the reliability of one of those
nonperturbative methods that have been used with some frequency in the
literature: The optimized perturbation theory  (OPT).  The OPT is an
analytical technique which allies the computational advantages of
ordinary  perturbation theory to a variational criterion in order to
generate nonperturbative  results~\cite{Okopinska:1987hp}.  The OPT
method has been used extensively in the literature to treat many
different physical systems, ranging from condensed matter problems,
phase transition problems in finite temperature quantum field theory
and others (see, e.g., 
Refs.~\cite{opt1,opt2,opt3,opt4,opt5,opt6,optaips,opt7,opt8,opt9,opt10,opt11,opt12,opt13,opt14,opt15,opt16} 
and references there in for some examples of applications).  

When applying the OPT method to a gauge theory, a modified form of the method is required.
In this case, a suitable modification preserving gauge invariance can be implemented. This modification
of the OPT method is known in this case as the Hard Thermal Loop perturbation theory (HTLpt)
(see, e.g., the original proposal in Ref.~\cite{Braaten:1991gm} and also the 
review~\cite{Andersen:2004fp}).

One particular issue regarding the OPT method that we would like to
also address in this work regards the quantity to which one should
apply the variational criterion as required  by the method. In a
calculation where different physical quantities are available, in the
original proposal by Stevenson~\cite{Stevenson:1981vj}, the
variational principle used was the principle of minimal sensitivity
(PMS) and it was advocated that the PMS  should be applied to each
different physical quantity that is being  computed, producing
different optimized parameters. However, one could argue that the PMS
should be applied to a more general quantity such as the ground-state
energy density as in Ref.~\cite{Krein:1995rp},  or to the effective
potential,  which generates all one-particle irreducible
contributions, as in Refs.~\cite{opt6,opt7,opt8,opt9,Kneur:2006ht},  while previous
works~\cite{Pinto:1999py,Pinto:1999pg} have shown that applying the
PMS to the self-energy would be more appropriate. In this work we
want to clarify in this issue of which quantity we should optimize in
the OPT method and also which quantity can provide the best
convergence in the OPT. With this aim we shall compare the results
obtained by a direct optimization of the effective potential (the
zero-point Green function),  the self-energy (the two-point Green
function) and also to the effective coupling (the four-point Green
function) in the context of the 0-dimensional $O(N)$ scalar
field theory model. 

One should note that one of the main differences as far as a comparison to
a quantum field theory model in  $D>0$ is concerned, is the need
to regularize and renormalize physical quantities, which is, of course, absent 
in the zero-dimensional model studied here.
The renormalization group flow dictates the change of physical parameters with the scale
and in this case the application of the OPT has to
be handled very carefully~\cite{Kneur:2013coa,Kneur:2015moa,Kneur:2015uha}. Note, however, that
the OPT method is not restricted to renormalizable models and it has been applied
successfully to many effective nonrenormalizable models as well~\cite{Krein:1995rp,Kneur:2010yv,Kneur:2012qp}.
Even so, the application of the OPT to the present exact soluble model offers an unique
opportunity to elucidate on the possible optimization criteria issues,
which are not possible to perform in other models without exact
solutions. Because of this, the 0-dimensional $O(N)$ scalar field theory model
is the perfect benchmark toy model to use to perform different tests related to the application
of the OPT method, but it also useful to test other different nonperturbative methods
used in quantum field theory as well. 

The remainder of this work is organized as follows. In Sec.~\ref{int},
we briefly describe the 0-dimensional $O(N)$ scalar field model and show why 
perturbation theory  might not be reliable in the context of this model in special. In Sec.~\ref{optsec}, we
introduce the OPT method and also describe three main variational
tools that are commonly used in conjunction with this method.  In
Sec.~\ref{comparison}, we perform a comparison between exact results
and the nonperturbative results obtained by OPT.  These results are
also contrasted with those obtained from the $1/N$-expansion. This way
we can better evaluate the  usefulness of the OPT and the
corresponding variational methods with this popular nonperturbative
method.  Our concluding remarks are given in
Sec.~\ref{conclusions}. Four appendixes are included where we give
some of the technical details.


\section{The 0-dimensional $O(N)$ scalar field model}
\label{int}

The 0-dimensional $O(N)$ scalar field model describes an
$N$-component anharmonic oscillator in zero spacetime dimension, whose
action is given by

\begin{eqnarray}
S (\boldsymbol{\varphi}) &=& \frac{m}{2}{\boldsymbol \varphi}.{\boldsymbol\varphi} +
\frac{\lambda}{4!}({\boldsymbol\varphi}.{\boldsymbol\varphi})^{2}, \label{action}
\end{eqnarray}
where $m, \lambda$ are real and positive parameters and ${\boldsymbol\varphi}
\equiv (\varphi_1, \ldots, \varphi_N)$ is a scalar
field with $N$ components. Equation~(\ref{action}) is an invariant under $O(N)$
rotations.

The generating function for the $n$-point Green's functions is given by

\begin{equation}
Z\left({\boldsymbol J}\right)= \int D\boldsymbol{\varphi}
\exp\left[{-S\left(\boldsymbol{\varphi}\right) +
  \boldsymbol{J}\cdot\boldsymbol{\varphi}}\right],
\label{fullger}
\end{equation}
where $\boldsymbol{J}$ is an external source. {}From the generating
function, the $n$-point Green functions are  given by

\begin{eqnarray}
G_{i_1 \cdots i_n}^{(n)} =\left.\frac{1}{Z}\frac{\delta^n Z({\boldsymbol J})}{\delta
   J_{i_1}\cdots \delta J_{i_n}}\right|_{J_i=0}= \langle\varphi_{i_1} \cdots
\varphi_{i_n}\rangle,
\end{eqnarray}
and averages are given by their standard definition, 

\begin{equation}
\langle\cdots\rangle =  \frac{\int \mathcal{D}\boldsymbol{\varphi}
  \exp[-S(\boldsymbol{\varphi})][\cdots]}{Z}.
\end{equation}
 The connected Green functions are obtained from
  the functional generator $W( \bf J ) $, defined as~\cite{ramond}

\begin{equation}
W (\boldsymbol{J})=\ln \left[\frac{Z(\boldsymbol{J}\ )}{Z_{0}}\right], 
\label{functionw}
\end{equation}
where $Z_0$ is the normalization of the generating functional
$Z_0=Z(\boldsymbol{J}=0)|_{\lambda=0}$.  The connected Green functions are
then given by

\begin{equation}
G_{c,{i_1} \ldots {i_n}}^{(n)}= \left.\frac{\delta^{n}
  W\left(\boldsymbol{J}\right)}{\delta J_{i_1} \ldots  \delta
  J_{i_n}}\right|_{ J_i=0} = \langle \varphi_{i_1} \ldots \varphi_{i_n}
\rangle_{\rm connected}. 
\label{green}
\end{equation}

{}From the definition of the expectation value of the field,

\begin{equation}
\phi_i=\langle \varphi_i\rangle_{J_i}=\frac{\delta
  W}{\delta J_i},
\end{equation}
we can perform the usual Legendre transformation and obtain the
effective action, or   the generation functional of the one-particle
irreducible (1PI) Green's functions~\cite{ramond}

\begin{equation}
\Gamma(\boldsymbol{\phi}) = W(\boldsymbol{J}) - \boldsymbol{J}\cdot \boldsymbol{\phi} .
 \label{functionG}
\end{equation}

The advantage of working with the 0-dimensional $O(N)$ scalar field model is that
it has explicit analytical solution, which can be contrasted with different
approximations used in the literature and, thus, it is a perfect benchmarking model
to use. In this work, we will make use of the effective potential $V_{\rm eff}$,
the self-energy $\Sigma$ and the 1PI four-point Green's functions. 
The one-particle irreducible (1PI) Green's functions,  $\Gamma^{(n)}$, are defined by
\begin{equation}
\Gamma^{(n)}_{i_1 \cdots i_n} = \left.-\frac{ \delta^n \Gamma[\boldsymbol{\phi}]}{\delta 
\phi_{i_1} ... \delta  \phi_{i_n}}\right|_{ \phi=0}
\end{equation}
In particular, in the rest of
this work we will make use of the effective potential, $V_{\rm
  eff}=-\ln Z$,  the self-energy,
$\Sigma \equiv \Gamma^{(2)} - m$ and the vertex function
$\Gamma^{(4)}$. 

Let us
give the explicit expressions for these quantities for the present model.
Starting with the partition function,
${Z} \equiv Z(0)$,  given by~\cite{Keitel:2011pn,hyperspherical}

\begin{equation}
Z = \int \mathcal{D}\boldsymbol{\varphi}\ e^{-S(\boldsymbol{\varphi})} =
\Omega_{N}\mathcal{R}_{N-1},
\label{partfunc}
\end{equation}
where $\Omega_{N}$ is the surface area in $N$-dimensional unit sphere, 

\begin{equation}
\Omega_N = \frac{2 \pi^{N/2}}{\Gamma\left(\frac{N}{2} \right)},
\end{equation}
and $\mathcal{R}_{N}$ is defined by

\begin{eqnarray}
\mathcal{R}_{N}&=&\int_{0}^{\infty }
\ x^{N}\ e^{-\frac{m}{2}x^{2}-\frac{\lambda}{4!}x^{4}}dx
\nonumber\\ &=&2^{\frac{3N-5}{4}}3^{\frac{N+1}{4}}\lambda^{-\frac{N+3}{4}}
\left[ \sqrt{\lambda}\Gamma \left(
  \frac{N+1}{4}\right)\right.\nonumber\\
  &\times&\left.\ _{1}F_{1}\left(\frac{N+1}{4};\frac{1}{2};
\frac{3m^{2}}{2\lambda}\right)-\sqrt{6}m\Gamma
  \left(  \frac{N+3}{4}\right)\right.\nonumber\\
&\times&\left.  \ _{1}F_{1}\left(
  \frac{N+3}{4};\frac{3}{2};\frac{3m^{2}}{ 2\lambda}\right) \right],
\label{RN-1}
\end{eqnarray}
where $_{1}F_{1}\left(\alpha;\beta;z\right)$ is the Kummer confluent
hypergeometric function~\cite{stegun}.

The explicit exact solutions for $V_{\rm eff}$, 
$\Sigma$ and $\Gamma^{(4)}$
that we use throughout this work are then found to be given, 
respectively, by:

\begin{equation}
V_{\rm{eff}}^{\rm exact} = -\ln(\Omega_N\mathcal{R}_{N-1}),
\label{Vexact}
\end{equation}

\begin{equation}
\Sigma_{\rm exact}=N\frac{\mathcal{R}_{N-1}}{\mathcal{R}_{N+1}} - m,
\label{Sigmaexact}
\end{equation}

\begin{eqnarray}
\Gamma^{\left( 4\right)}_{\rm exact} \!=\! 
- 3N^2 \!\left(\frac{\mathcal{R}_{N-1}}{\mathcal{R}_{N+1}}\right)^2\!\!
\left[\frac{N (\mathcal{R}_{N+3})\left(\mathcal{R}_{N-1}\right)}
{(N+2)\left(\mathcal{R}_{N+1}\right)^2} 
- 1 \right].
\label{Gammaexact}
\end{eqnarray}

\begin{figure}[ht]
\begin{center}
\includegraphics[width=0.9\linewidth,angle=0]{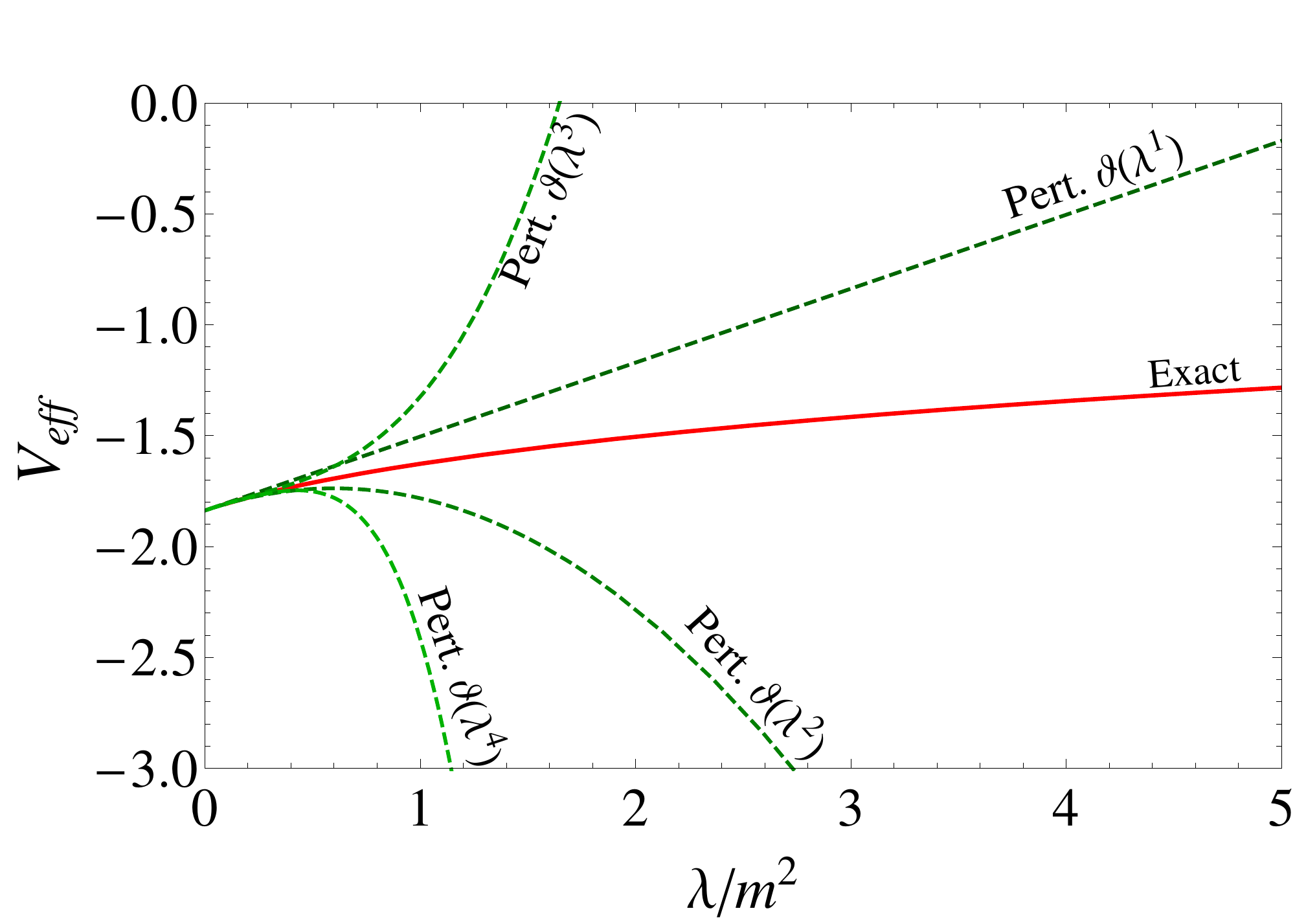}
\caption{The effective potential $V_{\rm eff}$ for the model,
  considering $N = 2$. The exact solution is given by the solid line.
  The results obtained by expanding $V_{\rm eff}$ in a power series in
  $\lambda/{m^2}$ are given by the dashed lines. }
\label{fig1}
\end{center}
\end{figure}


{}For illustrative purposes, we give in~\ref{perturbationtheory} the
perturbative expansion (the power series expansion in the coupling
constant $\lambda$) for the effective potential, $V_{\rm eff}$, the
self-energy $\Sigma$ and the vertex  function $\Gamma^{(4)}$. That
this is not a well posed perturbative series (as far convergence is
concerned) is shown in {}Fig.~\ref{fig1} for the particular case of
the perturbative expansion for the effective potential. We can see
that perturbative expansion shows no sign of converging.  In fact, it
can be shown that the series has a zero radius of
convergence~\cite{kleinert}.

\section{Optimized Perturbation Theory}
\label{optsec}

The application of the OPT method starts by implementing a linear
interpolation in the action,

\begin{equation}
S\rightarrow S_{\delta}=\left(1-\delta \right)S_{0}\left(\eta
\right)+\delta S , \label{opt}
\end{equation}
where $\delta$ is a fictitious expansion parameter, which is used only
for bookkeeping  purposes and set at the end equal to one. The
parameter $\eta$ is an arbitrary mass term, fixed through an
appropriate variational method. Some common ways of fixing this
parameter will be described below.  It is through this variational
method that nonperturbative effects are included through the OPT mass
parameter $\eta$. The OPT method has been successfully applied in
different scenarios (see, e.g Refs.~\cite{opt1,opt2,opt3,opt4,opt5,opt6,opt7,opt8,opt9,opt15,ldeqm} 
and
references there in). In this work, we apply this nonperturbative
method to evaluate the $n$-point 1PI Green's functions  of the
0-dimensional $O(N)$ scalar field model described in the previous
section.  We compare the OPT results with the exact solution that
this particular toy model gives. {}For comparison purposes we also
contrast the OPT results with those obtained through another
nonperturbative method, obtained with the $1/N$-expansion for the
model. 

Applying the interpolation (\ref{opt}) in Eq.~(\ref{action}) 
gives

\begin{eqnarray}
S(\boldsymbol{\varphi}) &=&  \frac{m}{2}\boldsymbol{\varphi}.\boldsymbol{\varphi} +
(1-\delta)\frac{\eta}{2}\boldsymbol{\varphi}.\boldsymbol{\varphi}+  \delta
\frac{\lambda}{4!}(\boldsymbol{\varphi}.\boldsymbol{\varphi})^{2} \nonumber\\ & =&
S_{0}(\boldsymbol{\varphi},\eta) + S_{\delta}(\boldsymbol{\varphi},\eta),
\label{Sinter}
\end{eqnarray}
where 

\begin{equation}
S_{0}(\boldsymbol{\varphi},\eta) =
\frac{m+\eta}{2}\boldsymbol{\varphi}.\boldsymbol{\varphi},
\end{equation}
and 

\begin{equation}
S_{\delta}(\boldsymbol{\varphi},\eta) = -\delta
\frac{\eta}{2}\boldsymbol{\varphi}.\boldsymbol{\varphi} + \delta
\frac{\lambda}{4!}(\boldsymbol{\varphi}.\boldsymbol{\varphi})^{2},
\label{newinter}
\end{equation}
which is considered as the modified interaction term in the OPT
method.

In the OPT method the bookkeeping parameter $\delta$ never appears in the free quadratic 
action $S_0(\varphi,\eta)$, it only appears in the modified interaction
action (see, e.g., Eq.~(\ref{newinter}) given above). Then, if we now perform an usual perturbation expansion in terms
of this modified interaction
term, we typically have to truncate the perturbative series to some order in $\delta$. The expressions
now depend explicitly on the parameter $\eta$ added by the method. Through an appropriate variational
method, this $\eta$ parameter is then fixed. Three of these optimization methods we will study below.
It is at this point that nonperturbative information is brought because $\eta$ will depend on the various
couplings of the theory.

The generating functional, using Eq.~(\ref{Sinter}), becomes

\begin{eqnarray}
Z =\int \mathcal{D}\boldsymbol{\varphi}\ e^{-
  S_{0}(\boldsymbol{\varphi},\eta)}\ e^{-S_{\delta}(\boldsymbol{\varphi})}. 
\label{zdelta}
\end{eqnarray}
The strategy to evaluate the effective potential $V_{\rm eff}=-\ln Z$,
$\Sigma$ and $\Gamma^{(4)}$  using OPT is very similar as we would do
when using perturbation theory.  Using the interaction term (\ref{newinter}),
we can compute the physical quantity of interest expanding the result up
to some order $k$ in
$\delta$. The procedure is immediate if we use the exact expressions
Eqs.~(\ref{Vexact}), (\ref{Sigmaexact}) and (\ref{Gammaexact}),
by making the substitutions in those expressions, 
$m \to m + (1-\delta)\eta$, $\lambda \to \delta \lambda$
and then expanding the respective expression up to the desired order in $\delta$. 
{}For example, the effective potential, evaluated up to order
$\delta^{2}$, is given by

\begin{eqnarray}
V_{\rm{eff}}& = &
-\ln\left[\frac{2^{N}\pi^{N/2}\ (m+\eta)^{-N/2}\  
\Gamma\left(1+\frac{N}{2}\right)}{N\ \Gamma
    \left(\frac{N}{2}\right)}\right]\delta^{0}\nonumber\\  
& + & N
\left[
  \frac{\lambda(2+N)-12\eta(m+\eta)}{24(m+\eta)^{2}}\right]\delta^{1}
\nonumber\\  &
- &\left\{ N
\left[\frac{36\eta^{2}(m+\eta)^{2}-12\lambda(2+N)
\eta(m+\eta)}{144(m+\eta)^{4}}\right]\right.\nonumber\\  &
+ & N
\left.\left[\frac{\lambda^{2}(6+5N+N^{2})}{144(m+\eta)^{4}}\right]
\right\}\delta^{2} +  \mathcal{O}(\delta^{3}).
\end{eqnarray}
Likewise, the self-energy in the OPT (up to order $\delta^{2}$) is
given by

\begin{eqnarray}
\Sigma & = & \eta \ \delta^{0} +
\left[\frac{\lambda(2+N)-6\eta(m+\eta)}{6 (m+\eta)}\right]
\delta^{1}\nonumber\\  & - & \left[
  \frac{\lambda(2+N)\left[\lambda(4+N)-6\eta(m+\eta)\right]}{36
    (m+\eta)^{3}}\right]\delta^{2}\nonumber\\  & + &
\mathcal{O}(\delta^{3}),
\end{eqnarray}
and $\Gamma^{\left(4\right)}$ (up to order $\delta^{2}$) is given by

\begin{equation}
\Gamma^{\left(4\right)}  =    \lambda \ \delta^{1} -
\left[\frac{\lambda^{2}\left(8+N\right)}{6(m+\eta)^{2}}\right]\delta^{2}
+  \mathcal{O}(\delta^{3}). 
\label{gamma4delta}
\end{equation}
High order terms for $V_{\rm{eff}}$, $\Sigma$  and
$\Gamma^{\left(4\right)}$ can be founded in~\ref{highorderopt}.
Note that these expressions expressed as a power series in
$\delta$ they depend explicitly on the OPT parameter $\eta$.
This parameter is fixed using an appropriate variational
principle, as we explain next.

\subsection{Optimization procedures}

If we would perform the expansion in $\delta$ to all orders, then
after taking the limit  $\delta \rightarrow 1$, of course the $\eta$
dependence of the quantities would exactly cancel.  However,   this
expansion to all orders is impracticable. In other words, we need to
eventually truncate the series  at some order $k$ in $\delta$. This
means that a $\eta$ dependence is left in the results and this
parameter need to be fixed somehow.  In this work, we will study three
possible optimization procedures used to fix $\eta$ in the OPT method:
The Principle of Minimal Sensitivity (PMS), the Fastest Apparent
Convergence (FAC) and, finally, the Turning Point (TP) method. The PMS
is based on a variational principle~\cite{Stevenson:1981vj}. If a
physical quantity $\Phi$  does not depend originally on $\eta$, we
must then determine the value of $\eta$ that makes this quantity
minimally sensitive  to it. This is the basis for the PMS method,
which is then determines $\eta$ by requiring that the quantity $\Phi$
evaluated to some order $k$ in $\delta$, $\Phi^{(k)}$, must satisfy

\begin{equation}
\left.\frac{d\Phi^{\left(k\right)}}{d\eta}\right|_{\eta=\bar{\eta},
  \delta=1}= 0. 
\label{PMS}
\end{equation}
The PMS then provides a new mass term $\bar\eta$ that depends on the
original parameters of the theory,  e.g., the coupling constants, thus
bringing in the nonperturbative results. We must emphasize that it is
not always guaranteed the  existence of nontrivial solution for the
PMS Eq.~(\ref{PMS}) and we need to verify this in each PMS
application. When it is the case that we cannot find a solution, then 
we need to make use of some other optimization procedure. {}For
example, in the FAC procedure~\cite{opt10,opt11}), we  require that the
$k_{\rm th}$-coefficient of the expansion in $\delta$ of a physical
quantity $\Phi$,

\begin{equation}
\Phi^{\left(k\right)} = \sum_{i=0}^{k} c_{i}\delta^{i},
\label{fac}
\end{equation}
to satisfy

\begin{equation}
\left.\left[\Phi^{\left(k\right)} - \Phi^{\left(k-1\right)}
  \right]\right|_{\delta=1} = 0. 
\label{FAC}
\end{equation}
This condition is, thus, equivalent to taking the $k_{\rm
  th}$-coefficient in Eq.~(\ref{fac}) equal to zero.  One should note
that it is not at all guaranteed that the condition given by either
Eq.~(\ref{PMS}) or by Eq.~(\ref{fac}) might have necessarily a
nontrivial solution. Then a third method can be used. 
As proposed in Ref.~\cite{kleinert2}, in the cases that neither of the
PMS or the FAC have a solution, then we can make use of the TP 
method. The  TP method is  defined by the condition~\cite{kleinert2}

\begin{equation}
\left.\frac{d^{2}\Phi^{\left(k\right)}}{d\eta^2}\right|
_{\eta=\bar{\eta}, \delta=1} = 0. 
\label{TP}
\end{equation}

Explicit expressions found for the optimum $\bar{\eta}$, for each one of the optimization
procedures describe above, are given in~\ref{appD}.

In the next section we will study the nonperturbative OPT results
applied to the model explained in  Sec.~\ref{int}. We make a
comparison of the OPT results with those obtained from an expansion in
the number of components for the field, i.e., the large-$N$ (LN)
expansion, which is explained in~\ref{sectionlargen} and we also
compare these results with those obtained from the exact solution for the model. 


\section{Results}
\label{comparison}

In this section we will present our results using the OPT when
evaluating  the effective potential $V_{\rm eff}$,  the self-energy
$\Sigma$ and the vertex  function $\Gamma^{(4)}$ for the 0-dimensional
scalar field model.  These results are also contrasted with the ones
obtained using the LN expansion, presented in~\ref{sectionlargen} 
(see also Ref.~\cite{Keitel:2011pn} for
details).

\begin{figure}[!htb]
  \includegraphics[width=0.9\linewidth]{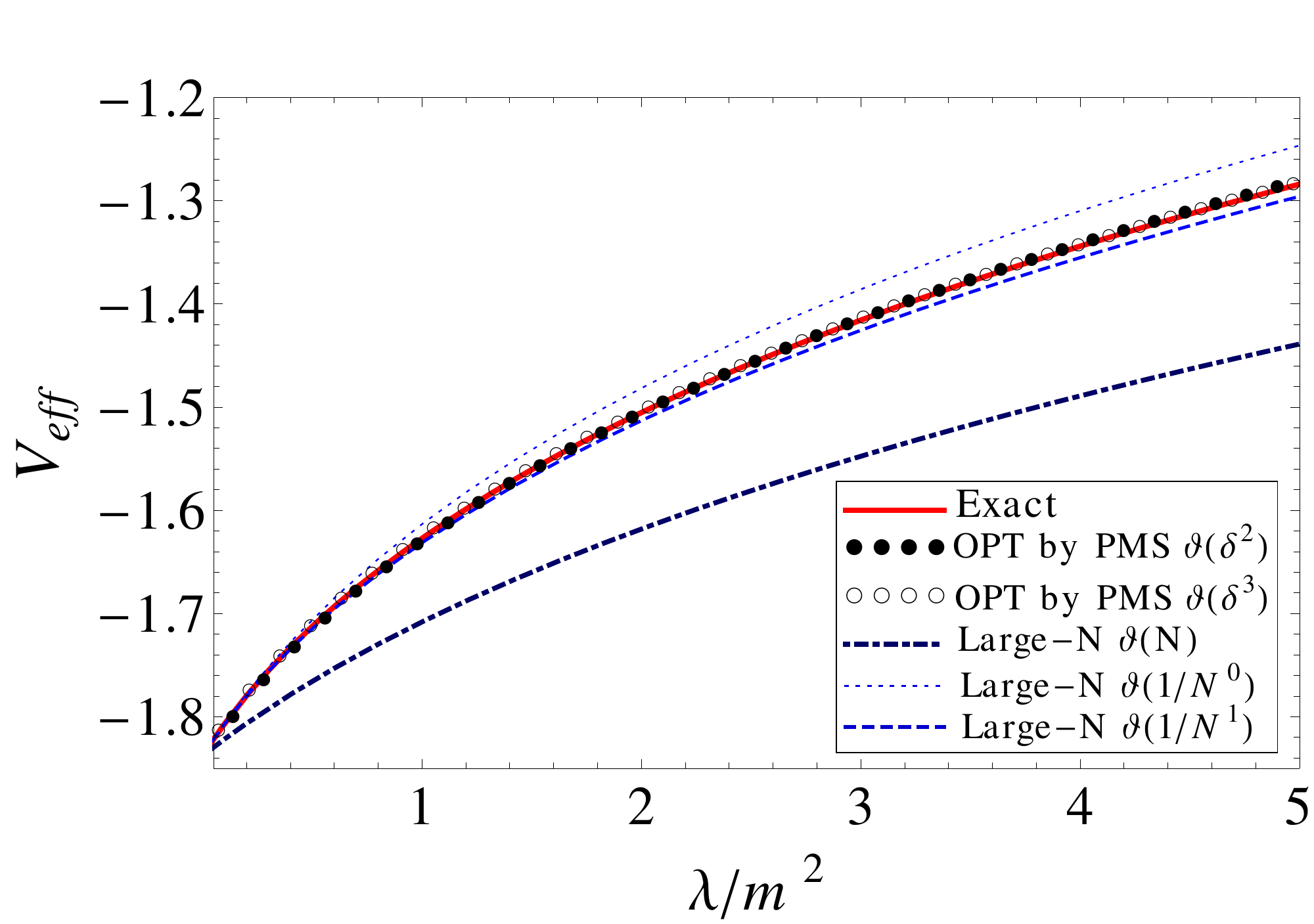}
 \caption{The effective potential $V_{\rm eff}$ for $N = 2$. The exact
   result (solid line),  the  LN results (dashed-dotted, dotted and
   dashed lines)  and the OPT results (circles),  where we optimize
   $\Sigma$  by using the PMS condition.}
\label{fig2}
\end{figure}
\begin{center}
\begin{figure*}
\subfigure[]{\includegraphics[width=6.75cm,height=6cm]{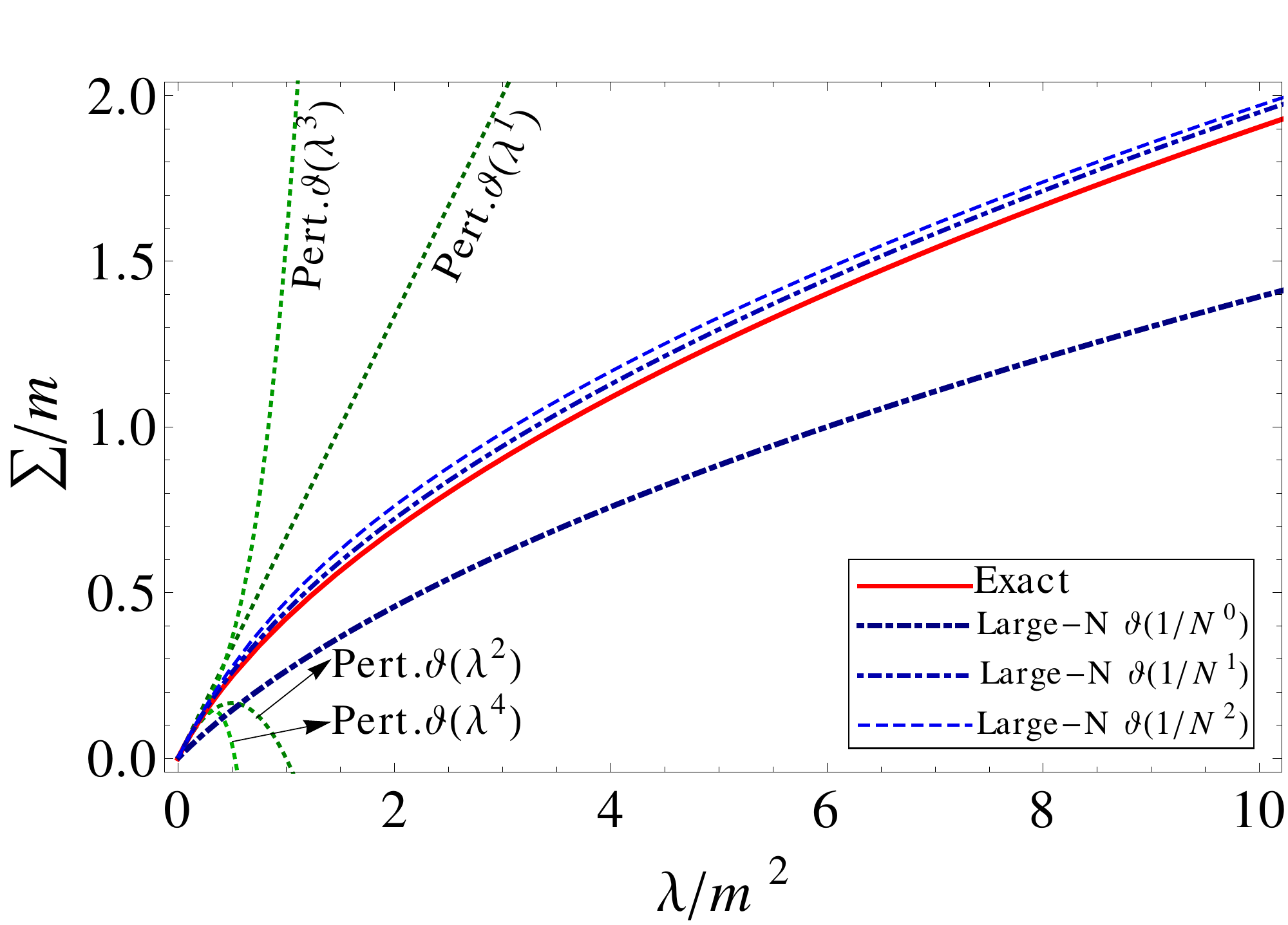}}\hspace{0.15cm}
\subfigure[]{\includegraphics[width=6.75cm,height=6cm]{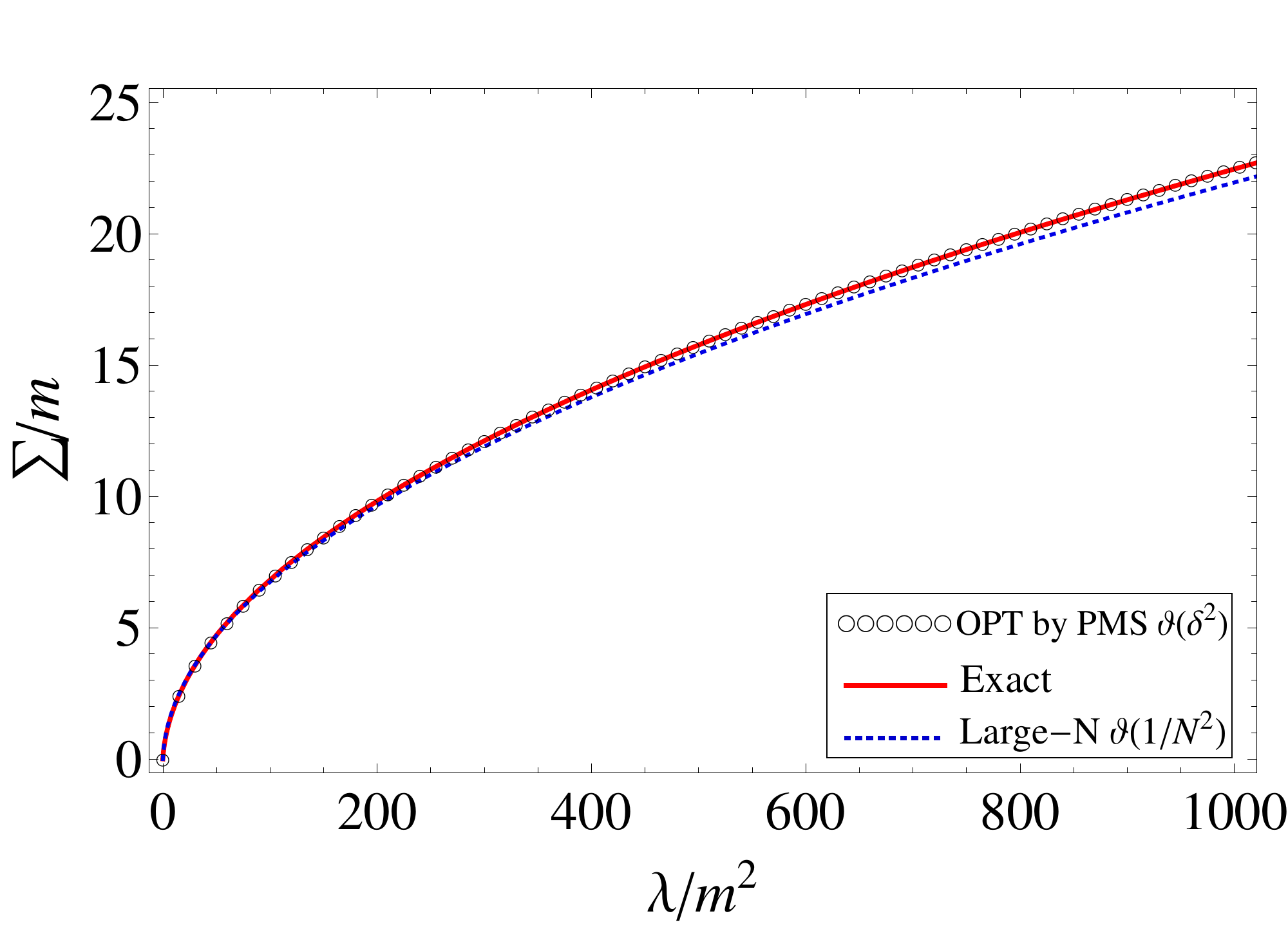}}
\caption{ The self-energy $\Sigma$ for $N=2$. Panel \textbf{(a)}: The
   exact result (solid line),  LN results (dotted and dashed-dotted
   lines) and the perturbative results (dotted lines) as a function of
   the  the coupling constant. Panel \textbf{(b)}: Extrapolation of
   the results for the strong coupling regime, for the cases of LN
   result, shown to $\mathcal{O}(\frac{1}{N^2})$ (dashed line), exact
   result (solid line) and  OPT result (circle), shown to
   $\mathcal{O}(\delta^2)$  optimizing $\Sigma$ by PMS.
}
\label{fig3}
\end{figure*}
\end{center}

In {}Fig.~\ref{fig2} we show the results for the effective potential
$V_{\rm eff}$ at $N=2$ in the cases of both the OPT and LN. Contrary
to the results obtained in perturbation theory and shown in
{}Fig.~\ref{fig1}, we see from the results in {}Fig.~\ref{fig2} that
the OPT and LN  both produce results with better convergence
properties, with the OPT at order $\delta^2$  already agreeing quite
well with the exact result. This agreement remains even when the
coupling is much larger, while the LN results, at increasing orders in
$1/N$, tend to oscillate around the exact solution.

Results for self-energy $\Sigma$ at $N=2$ are presented in
{}Fig.~\ref{fig3}. In the panel (a) of {}Fig.~\ref{fig3} we can again
see the bad behavior of perturbation theory.  In this same panel, we
also show the results for the LN for this case, while in the panel (b)
of {}Fig.~\ref{fig3} we show both the LN result at order $1/N^2$ and
the OPT at analogous order, $\delta^2$. Again we see that the OPT
covers much better the exact result even for very large values for the
coupling constant.  Here we have optimized $\Sigma$ by PMS (note that
PMS and TP do not present nontrivial solutions at
$\mathcal{O}(\delta^1)$, while the results from FAC are slight worser
than the ones obtained with the PMS).

\begin{center}
\begin{figure*}
\subfigure[]{\includegraphics[width=6.75cm,height=6cm]{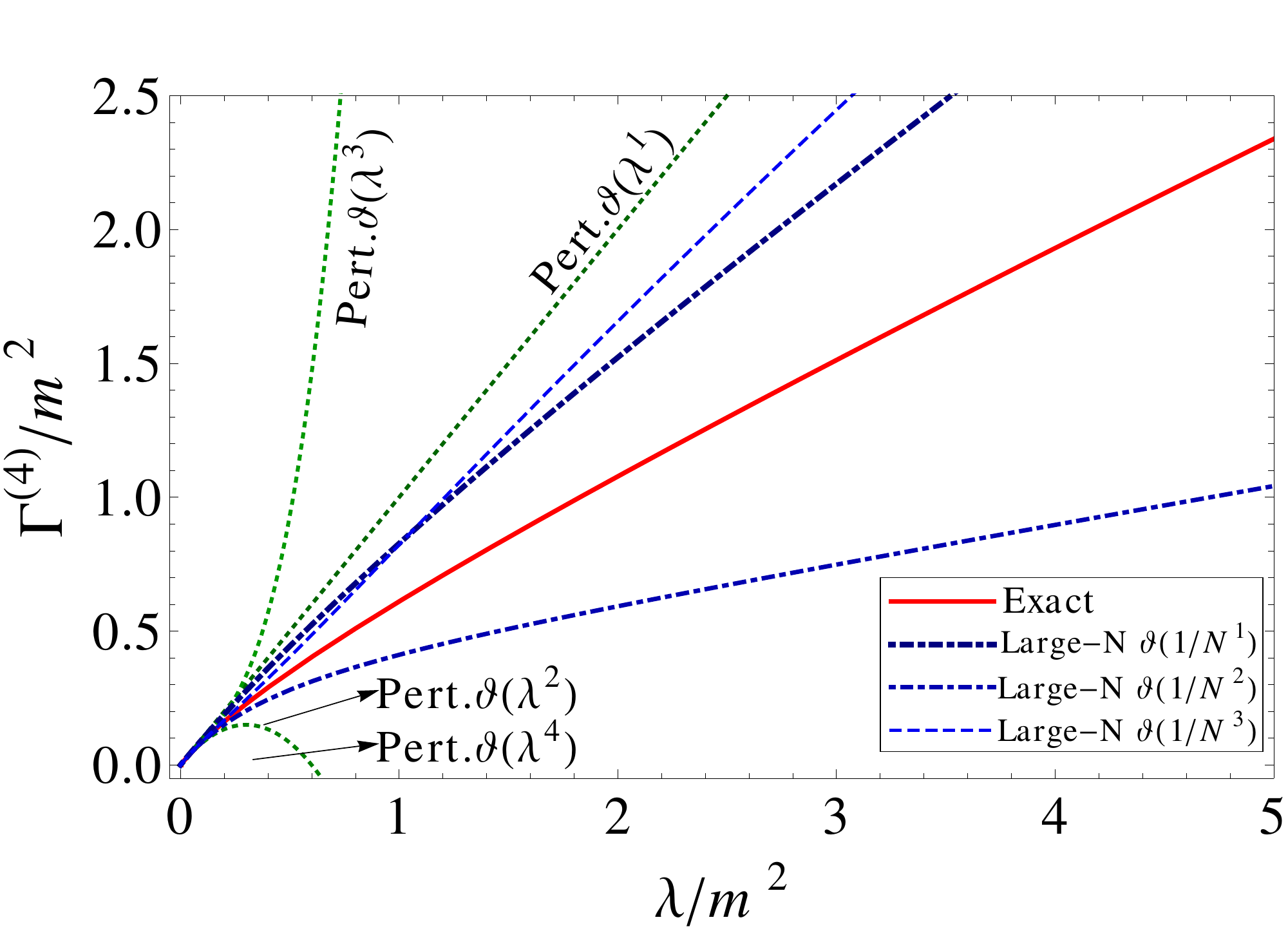}}\hspace{0.15cm}
\subfigure[]{\includegraphics[width=6.75cm,height=6cm]{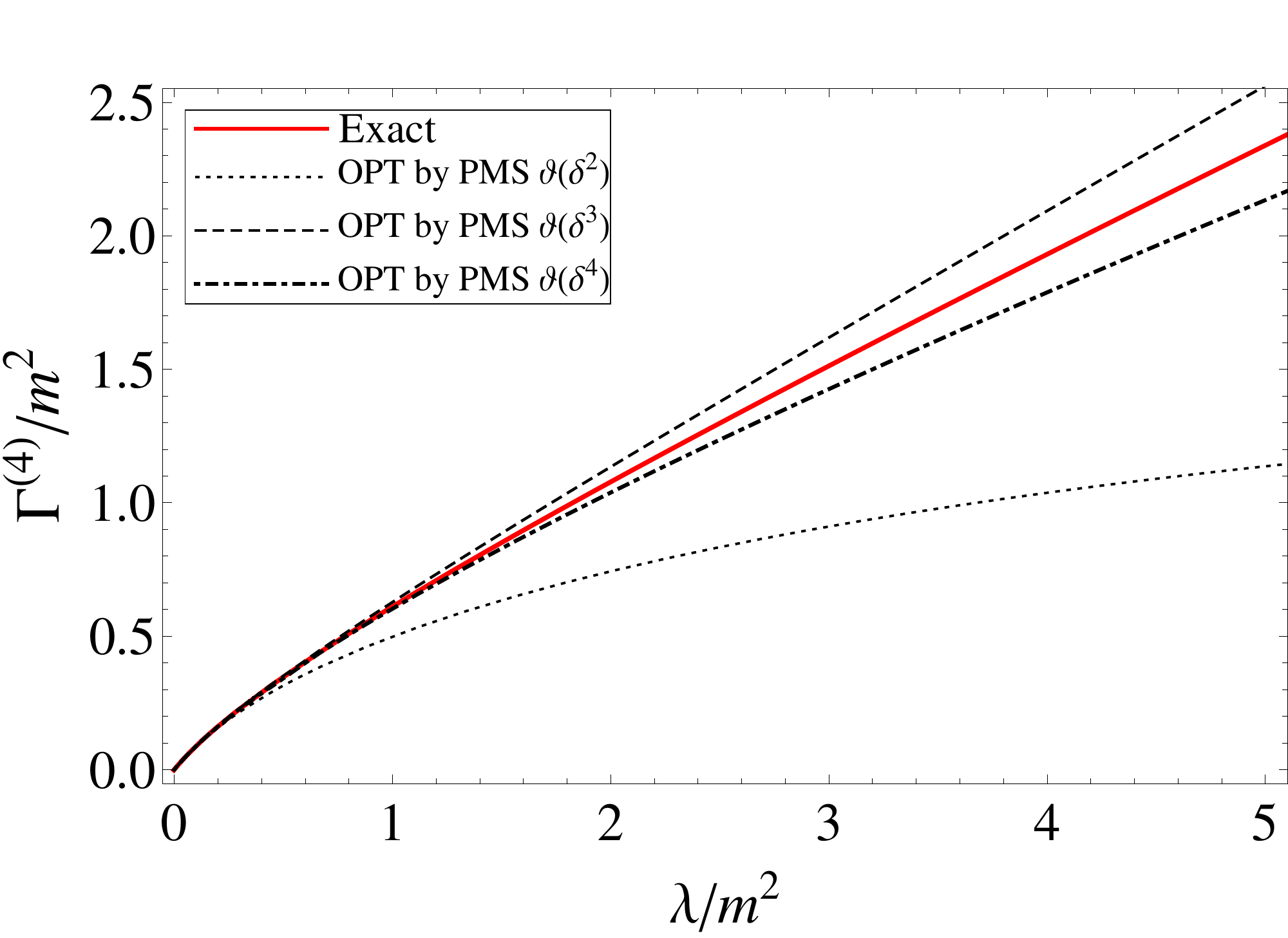}}
\caption{ Results for $\Gamma^{(4)}$ for $N = 2$. Panel \textbf{(a)}:
  the perturbative  results (dotted lines),  the LN results (dashed
  and dashed-dotted lines) and the exact solution (solid line). Panel
  \textbf{(b)}: The exact solution (solid line)  and the OPT results
  shown up to $\mathcal{O}(\delta^{4})$, which were obtained by
  optimizing $\Sigma$ using the PMS scheme.
}
\label{fig4}
\end{figure*}
\end{center}

Results for $\Gamma^{\left(4\right)}$ are presented in
{}Fig.~\ref{fig4}. In the panel (a) of {}Fig.~\ref{fig4} we show once
again the breakdown of perturbation theory, while the LN results
present strong deviations from the exact result.  In the panel (b) of
{}Fig.~\ref{fig4} we show how by increasing the order in the OPT it
oscillates around the exact solution, converging to it. In this case
we optimize $\Sigma$ by  PMS and use the solution in
$\Gamma^{\left(4\right)}$. The other optimization schemes, FAC and TP
produce results that are worse. Choosing to optimize the effective
potential, or the own function  $\Gamma^{\left(4\right)}$  also lead
to results that worse than optimizing the self-energy and using the
result back in $\Gamma^{\left(4\right)}$.  This shows that for this
case, optimizing the self-energy as a basic quantity, in the PMS
scheme, is the better choice.

\begin{center}
\begin{figure*}
\subfigure[]{\includegraphics[width=6.75cm,height=6cm]{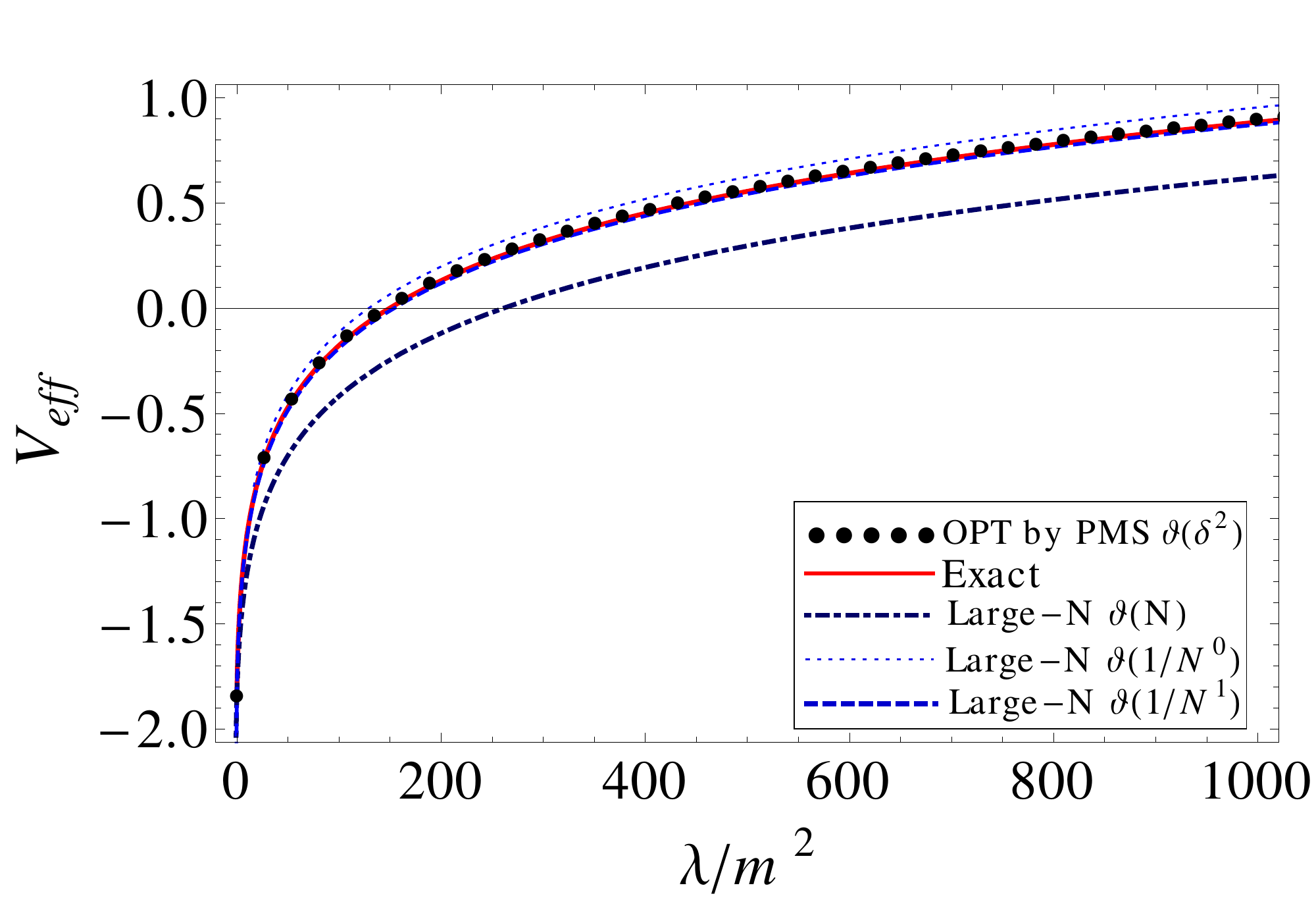}}\hspace{0.15cm}
\subfigure[]{\includegraphics[width=6.75cm,height=6cm]{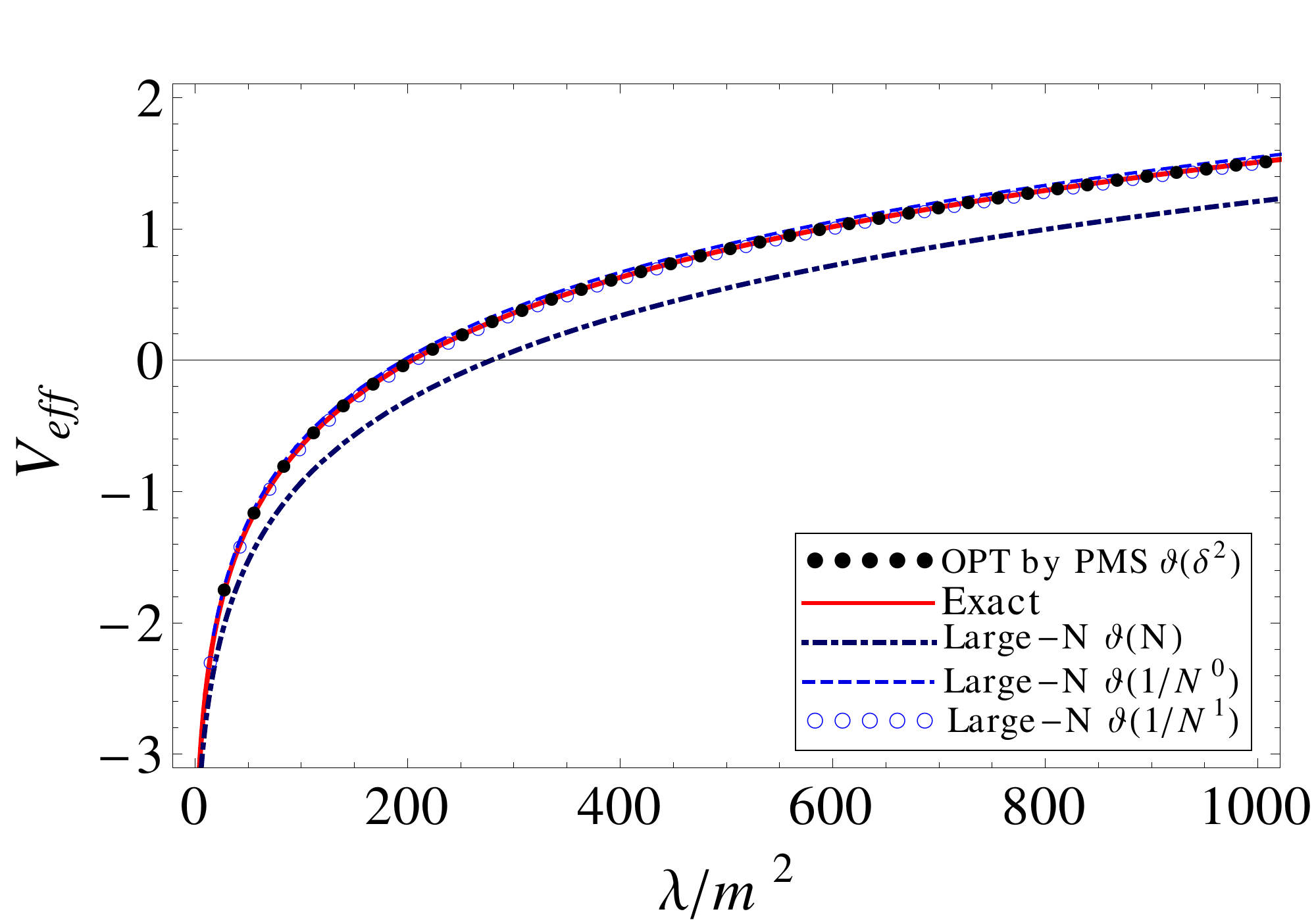}}\\
\subfigure[]{\centerline{\includegraphics[width=6.75cm,height=6cm]{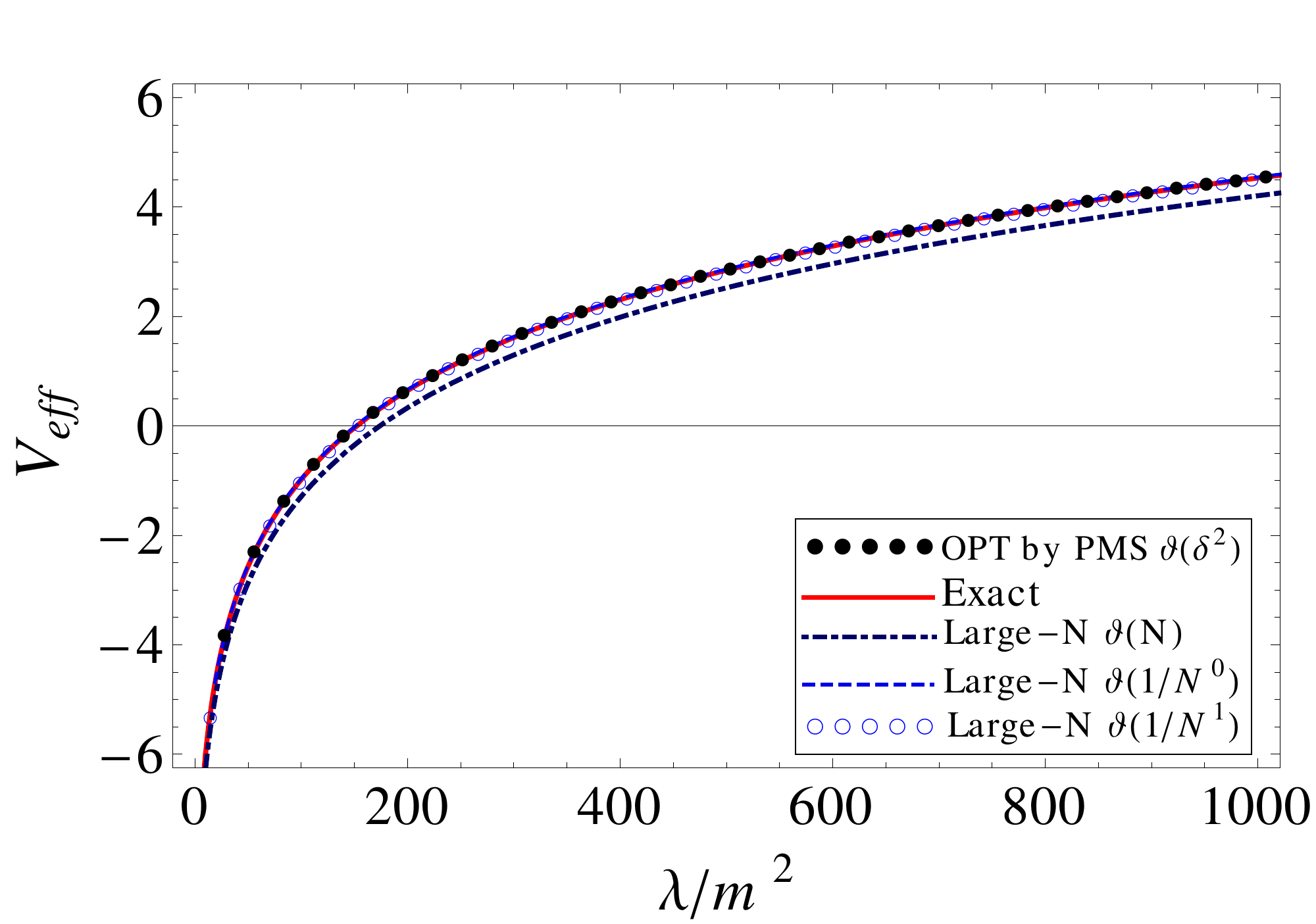}}}
\caption{ Results for the effective potential $V_{\rm eff}$ for the
   cases of $N=2$  (panel \textbf{(a)}), $N=4$ (panel  \textbf{(b)})
   and for $N=10$ (panel \textbf{(c)}). The exact solution is shown
   with a solid line, the LN  results are given by the dashed-dotted, dashed and 
   unfilled circles and the OPT results are represented by filled circles,
   shown up to $\mathcal{O}(\delta^2)$. The choice here was to
   optimize  $\Sigma$ using the PMS scheme.
}
\label{fig5}
\end{figure*}
\end{center}

In {}Fig.~\ref{fig5} we verify how the dependence on $N$ influences
the results for the effective potential.  Three cases are considered,
$N=2$, $N=4$ and $N=10$ and both OPT and LN are contrasted with the
exact solution.  In this case, as we have adopted in
{}Fig.~\ref{fig2}, we have chosen to optimize $\Sigma$ using the PMS
so to get the optimum  solution $\bar{\eta}$ and this solution is then
used back in $V_{\rm eff}$. Once again, the PMS applied on the
self-energy  is found to be the best optimization scheme.  The results
presented in  {}Fig.~\ref{fig5} also show that in general the OPT
presents more robust results than the LN approximation, with the OPT
converging faster to the exact result.

\begin{center}
\begin{figure*}
\subfigure[]{\includegraphics[width=6.75cm,height=6cm]{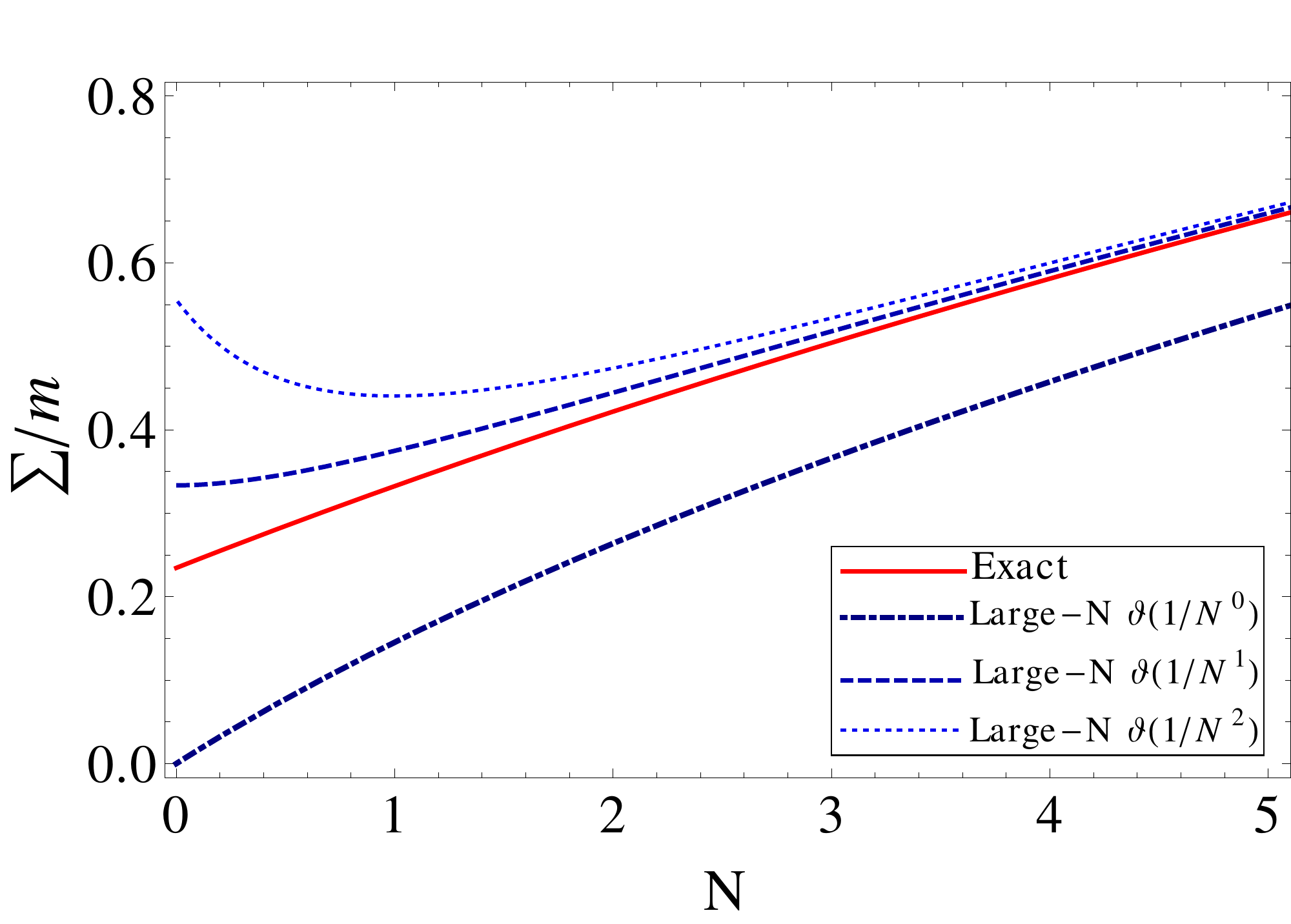}}\hspace{0.15cm}
\subfigure[]{\includegraphics[width=6.75cm,height=6cm]{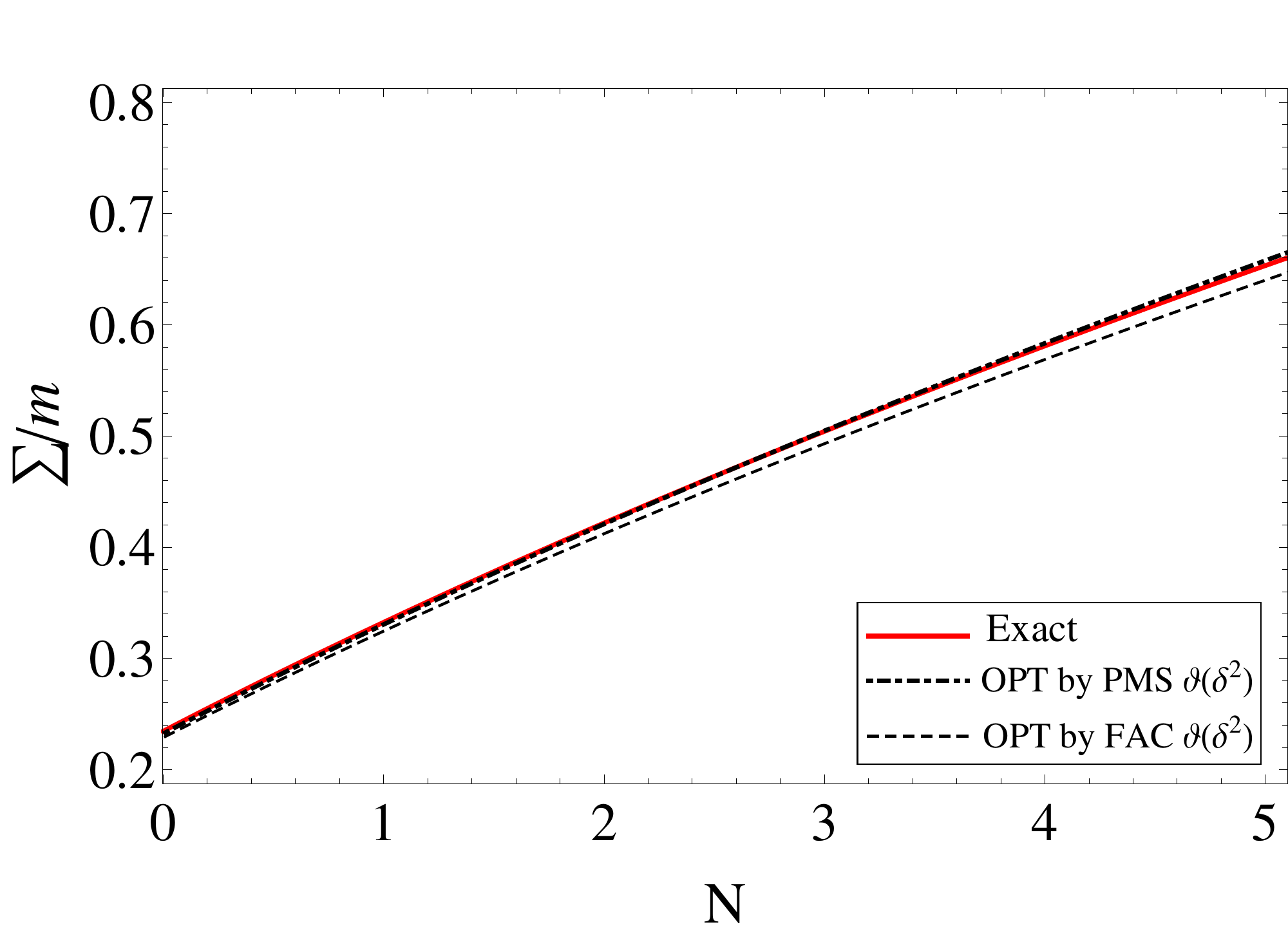}}
\caption{ Results for $\Sigma$, for a fixed value of
   $\frac{\lambda}{m^{2}}=1$, as function of $N$.  Panel \textbf{(a)}:
   The exact solution (solid line) and the LN results.  panel
   \textbf{(b)}: The exact solution (solid line) and the OPT results
   to $\mathcal{O}\left(\delta^{2}\right)$ and
   $\mathcal{O}\left(\delta^{3}\right)$, obtained by optimizing
   $\Sigma$ by PMS (dotted-dashed) and by FAC (dashed).
}
\label{fig6}
\end{figure*}
\end{center}

\begin{center}
\begin{figure*}
\subfigure[]{\includegraphics[width=6.75cm,height=6cm]{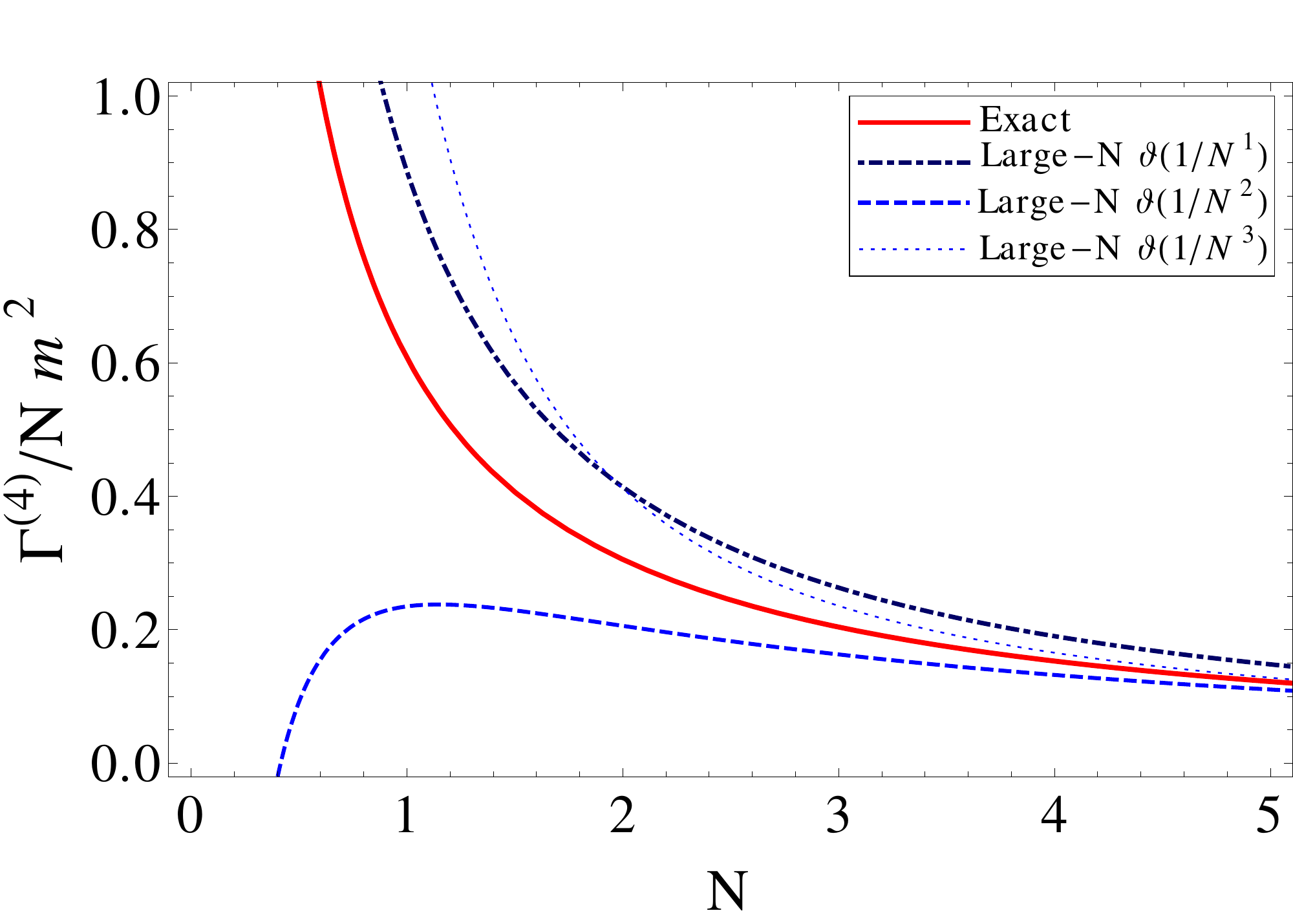}}\hspace{0.15cm}
\subfigure[]{\includegraphics[width=6.75cm,height=6cm]{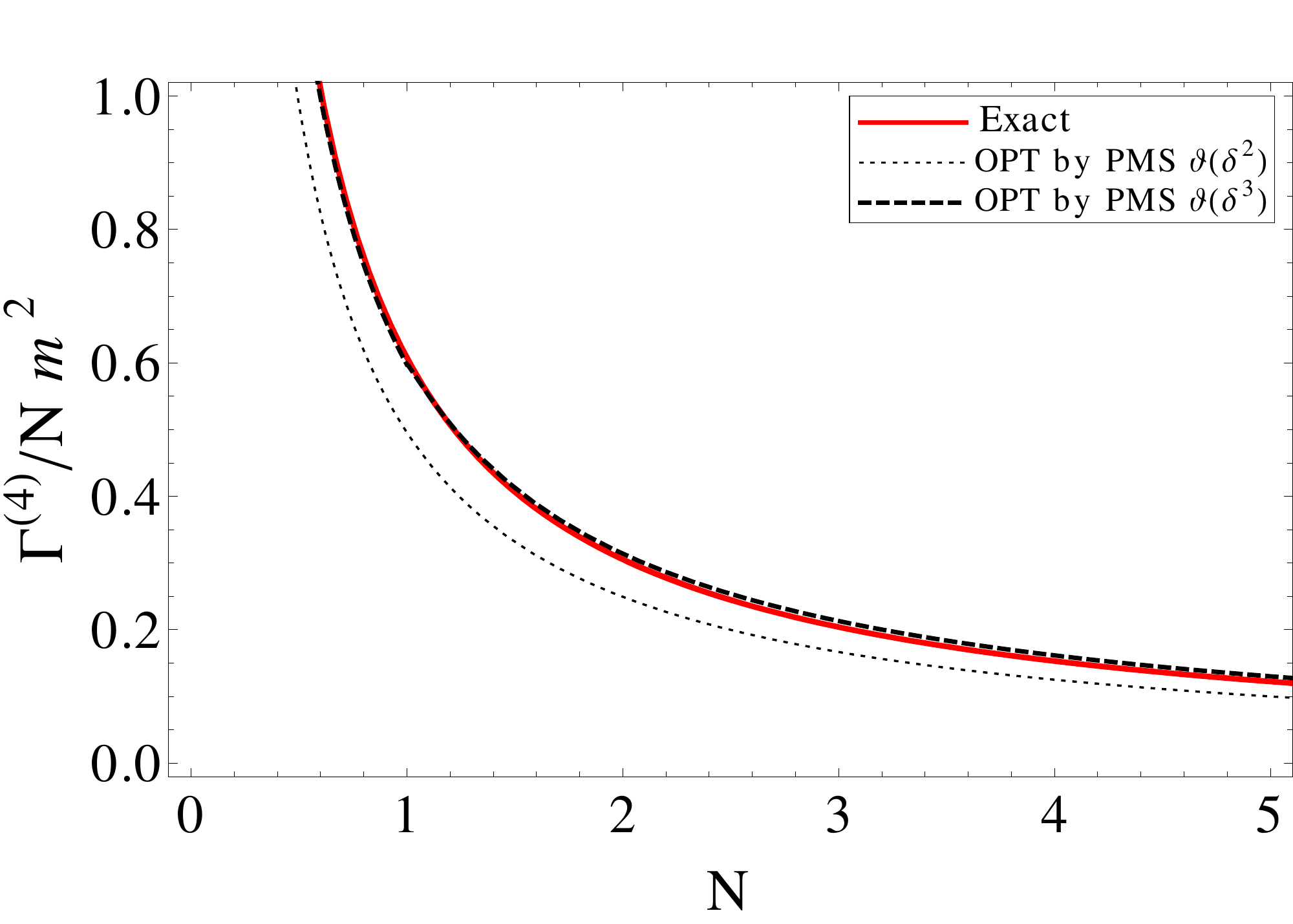}}
\caption{ Results for the four point function $\Gamma^{(4)}$, at the
   fixed value $\frac{\lambda}{m^{2}}=1$, as function of $N$.  Panel
   \textbf{(a)}: The exact solution (solid line) and the LN results
   (dashed, dashed-dotted and dotted lines).  Panel \textbf{(b)}: The
   OPT results to $\mathcal{O}\left(\delta^{2}\right)$ and
   $\mathcal{O}\left(\delta^{3}\right)$, obtained by optimizing
   $\Sigma$ by PMS.
}
\label{fig7}
\end{figure*}
\end{center}

In {}Figs.~\ref{fig6} and \ref{fig7} we also study the dependence of
the different methods as a function of the number of components $N$,
for the fixed value of the parameters in the model, $\lambda/m^2=1$,
for the self-energy $\Sigma$ and coupling function $\Gamma^{(4)}$,
respectively.  In {}Fig.~\ref{fig6}, where we show the self-energy
$\Sigma$ as a function of $N$,  we can see that the LN results tend to
present convergence to the exact solution only for very large values
of $N$.  As in the OPT case, when optimizing the own self-energy
$\Sigma$ using the  PMS scheme, it shows results that are much weaker
dependent on $N$, as far the convergence is concerned. The result from
the OPT are also much closer to the exact solution.  {}For comparison
purposes, we also show in the panel (b) of {}Fig.~\ref{fig6} the
results obtained using the FAC  optimization procedure (applied on
$\Sigma$), which also presents a very small deviation from the exact
solution, even for larger values of $N$, but it slight under estimates
the exact solution when compared with the PMS optimization procedure.

Likewise, in {}Fig.~\ref{fig7} we show the results for $\Gamma^{(4)}$
as a function of $N$ for the different approximation methods. In the
panel (a) of {}Fig.~\ref{fig7} we give the LN results, while in the
panel (b) we give the OPT results.  We note that LN results present a
strong deviation from the exact solution, only tending to converge (in
an oscillatory manner) at very larger values of $N$, while the OPT
results are robust at any order in $N$ and presents very good with the
exact solution already at $\mathcal{O}\left(\delta^3\right)$.

\begin{center}
\begin{figure*}
\subfigure[]{\includegraphics[width=6.75cm,height=6cm]{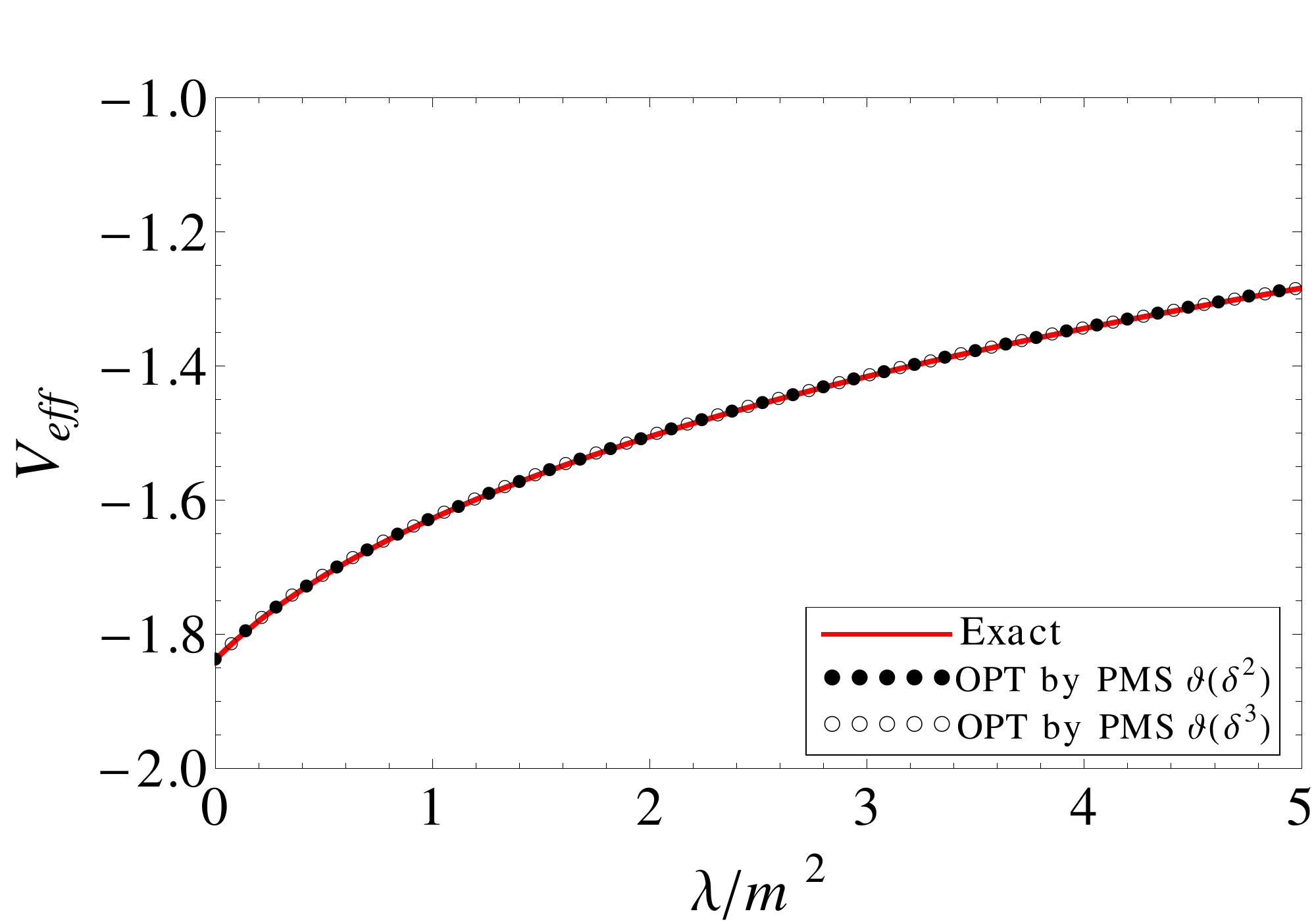}}\hspace{0.15cm}
\subfigure[]{\includegraphics[width=6.75cm,height=6cm]{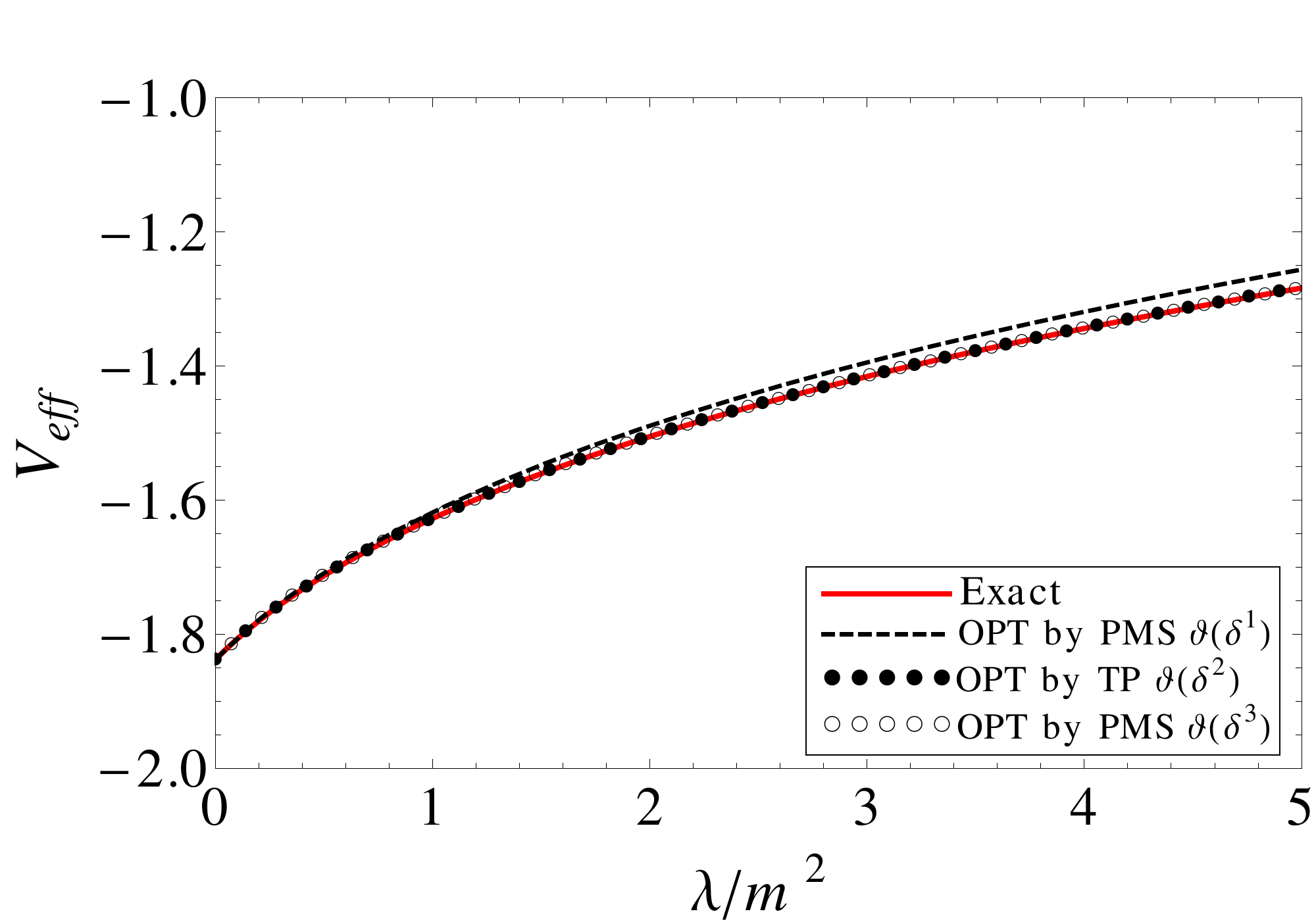}}\\
\subfigure[]{\includegraphics[width=6.75cm,height=6cm]{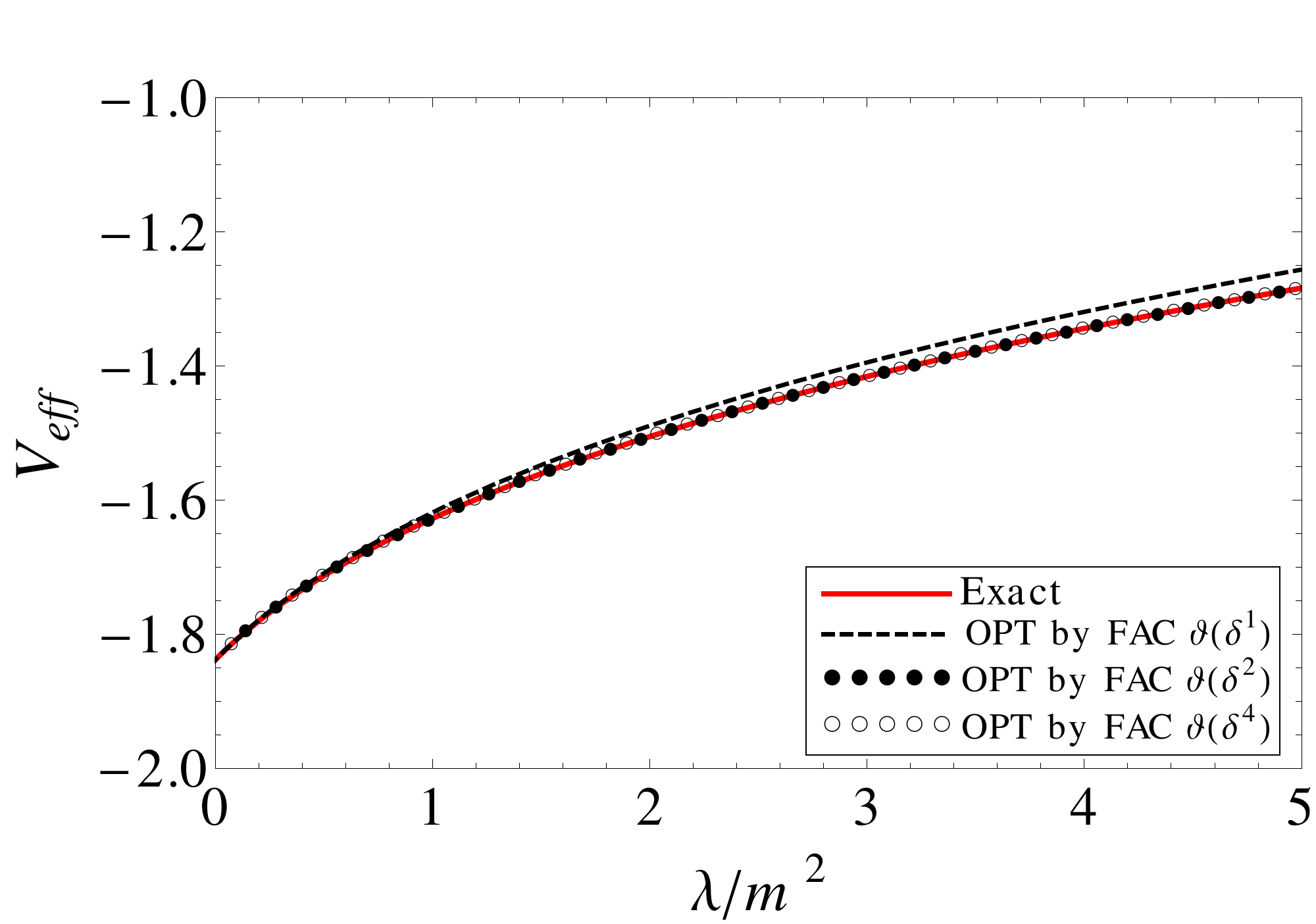}}\hspace{0.15cm}
\subfigure[]{\includegraphics[width=6.75cm,height=6cm]{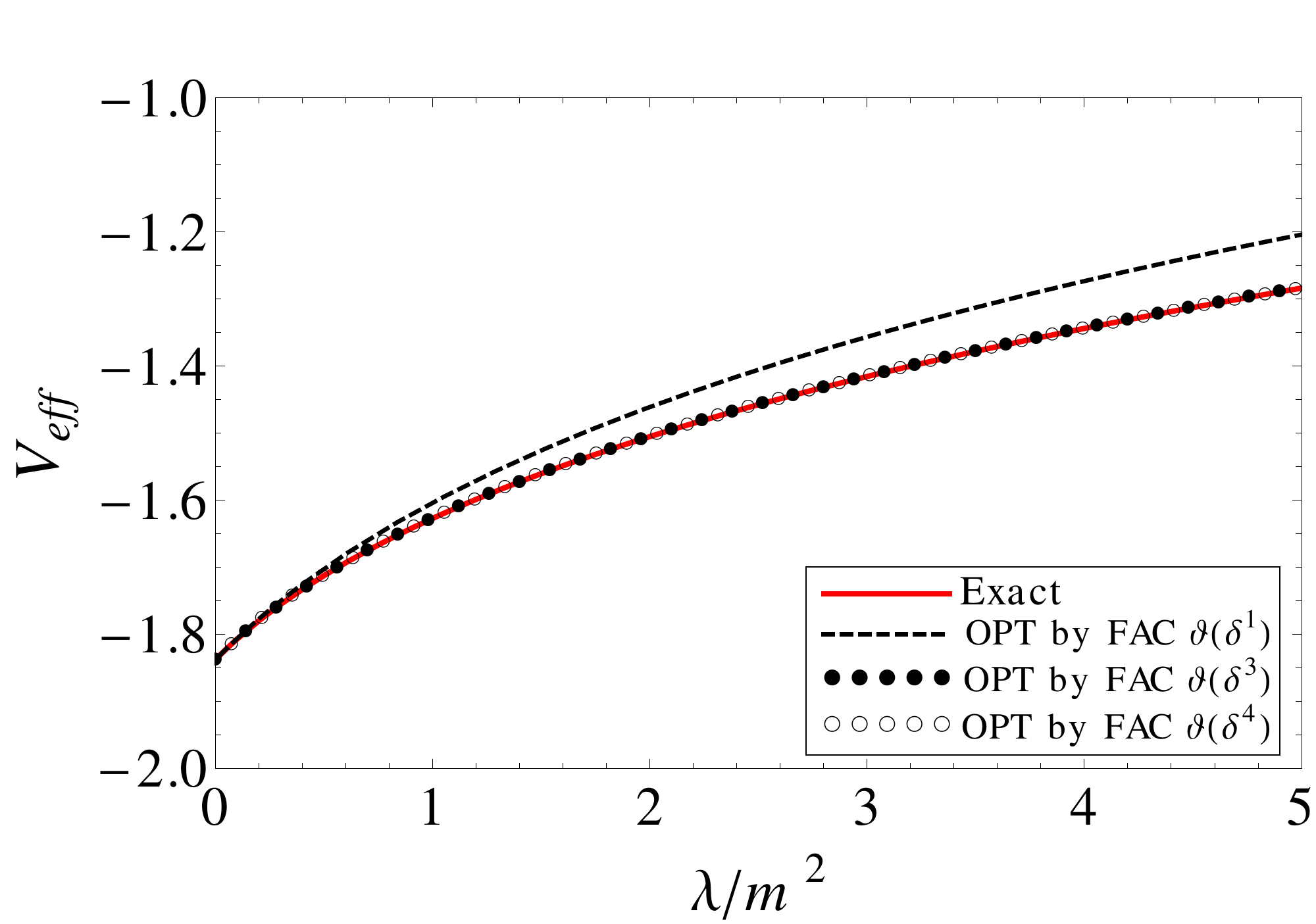}}
\caption{ Exact result (solid line) and OPT results (dashed line and
   circles) for $V_{\rm eff} $  at $N = 2$ when we optimize:  $\Sigma$
   by PMS (panel (a)),  $V_{\rm eff}$  by PMS and TP (panel (b)),
   $\Sigma$ by FAC (panel (c)) and  $V_{\rm eff}$  by FAC (panel
   (d)).
}
\label{fig8}
\end{figure*}
\end{center}

\begin{center}
\begin{figure*}
\subfigure[]{\includegraphics[width=6.75cm,height=6cm]{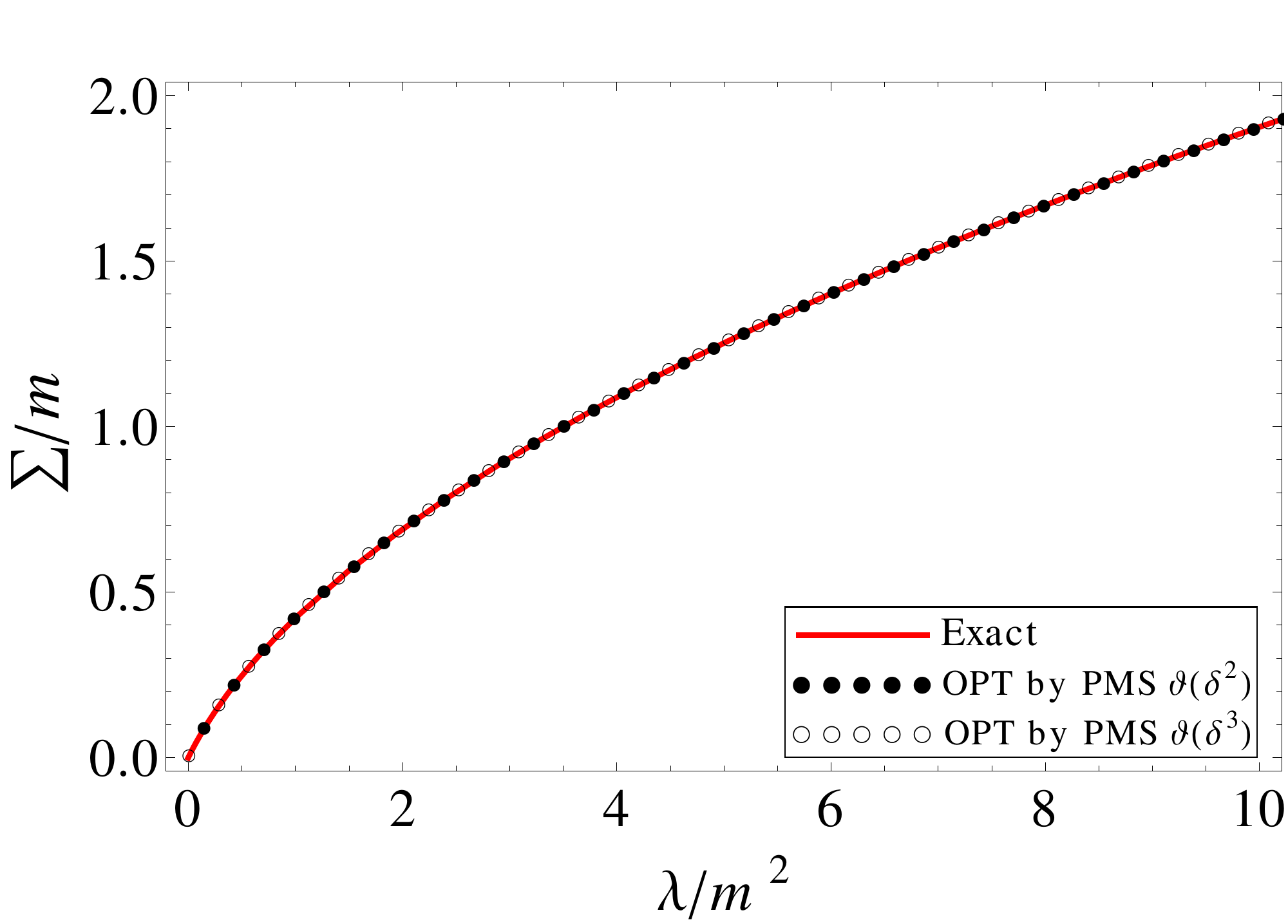}}\hspace{0.15cm}
\subfigure[]{\includegraphics[width=6.75cm,height=6cm]{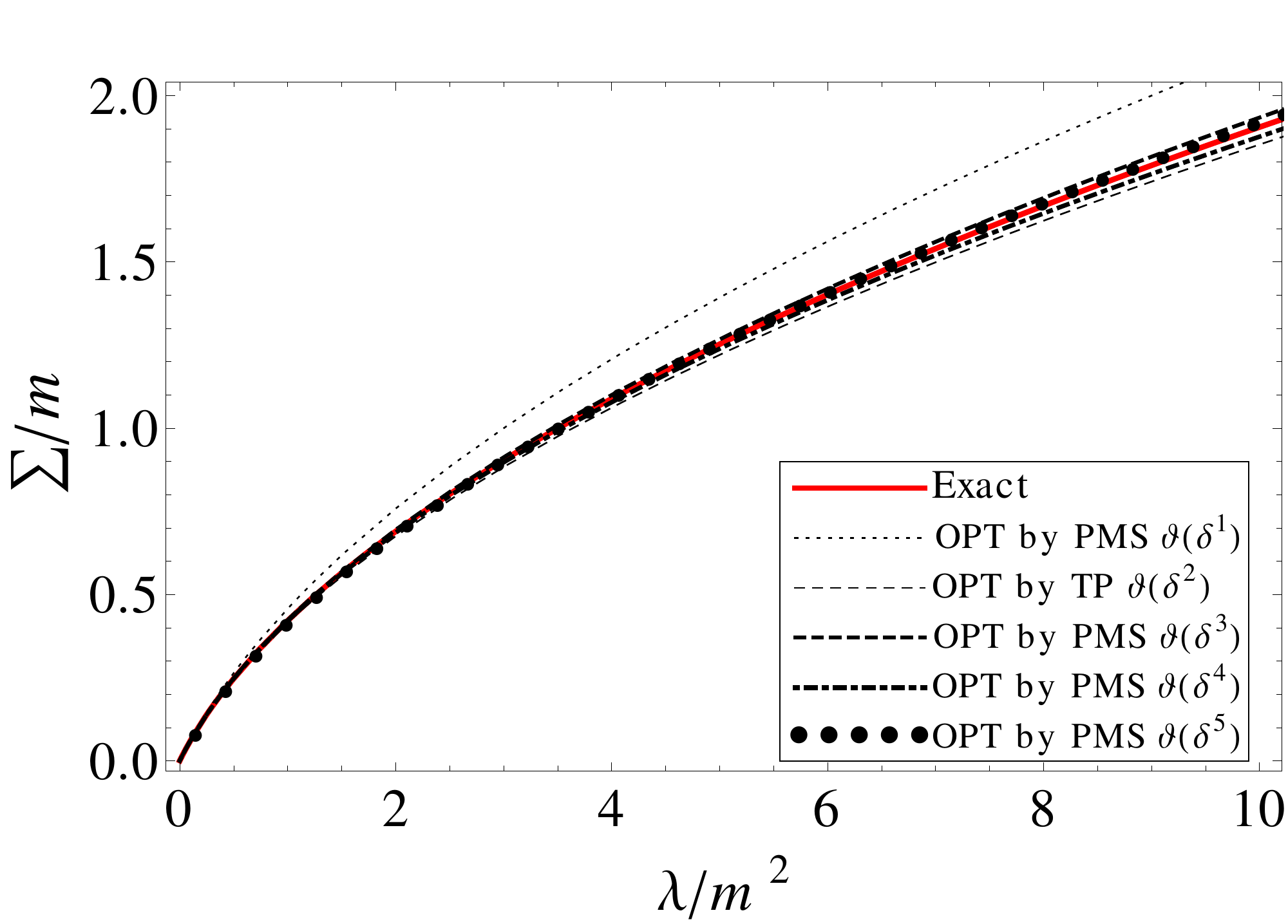}}\\
\subfigure[]{\includegraphics[width=6.75cm,height=6cm]{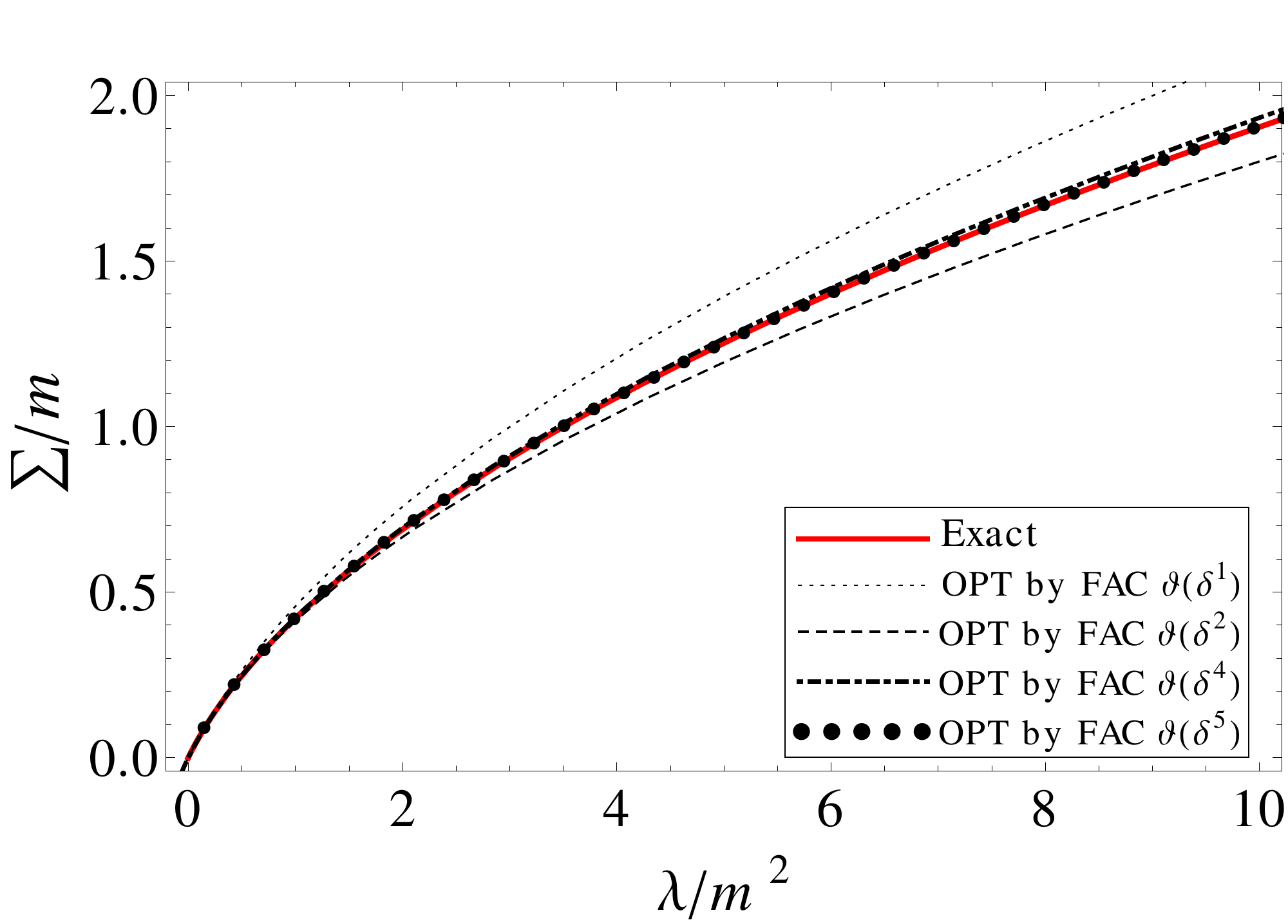}}\hspace{0.15cm}
\subfigure[]{\includegraphics[width=6.75cm,height=6cm]{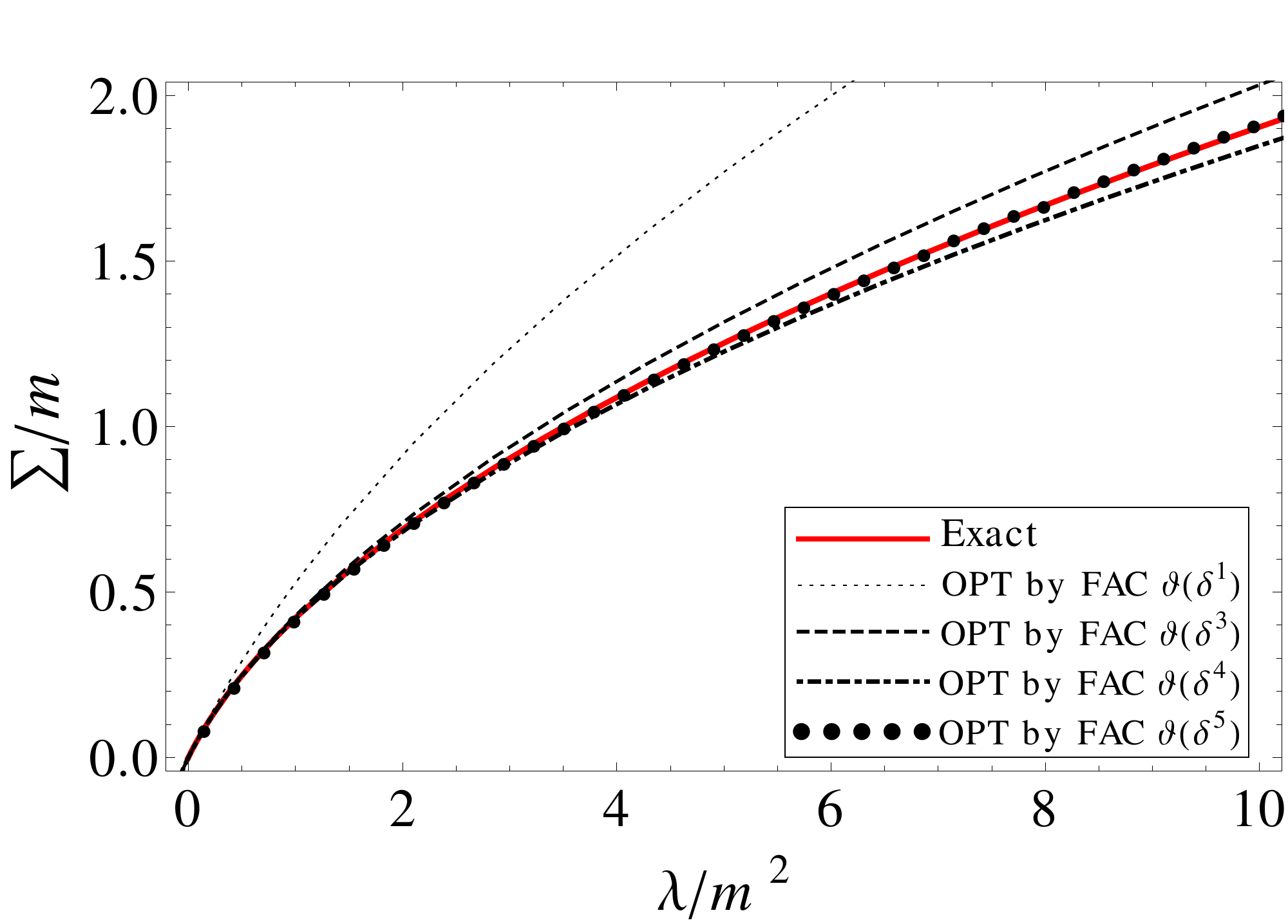}}
\caption{ Exact result (solid line) and OPT results for $\Sigma$ at
   $N=2$ when we optimize:  $\Sigma$ by PMS (panel (a)),  $V_{\rm
     eff}$ by PMS and TP (panel (b)),  $\Sigma$ by FAC (panel (c)) and
   $V_{\rm eff}$ by FAC (panel (d)).}
\label{fig9}
\end{figure*}
\end{center}

\begin{figure}[ht]
\begin{center}
\includegraphics[width=0.9\linewidth,angle=0]{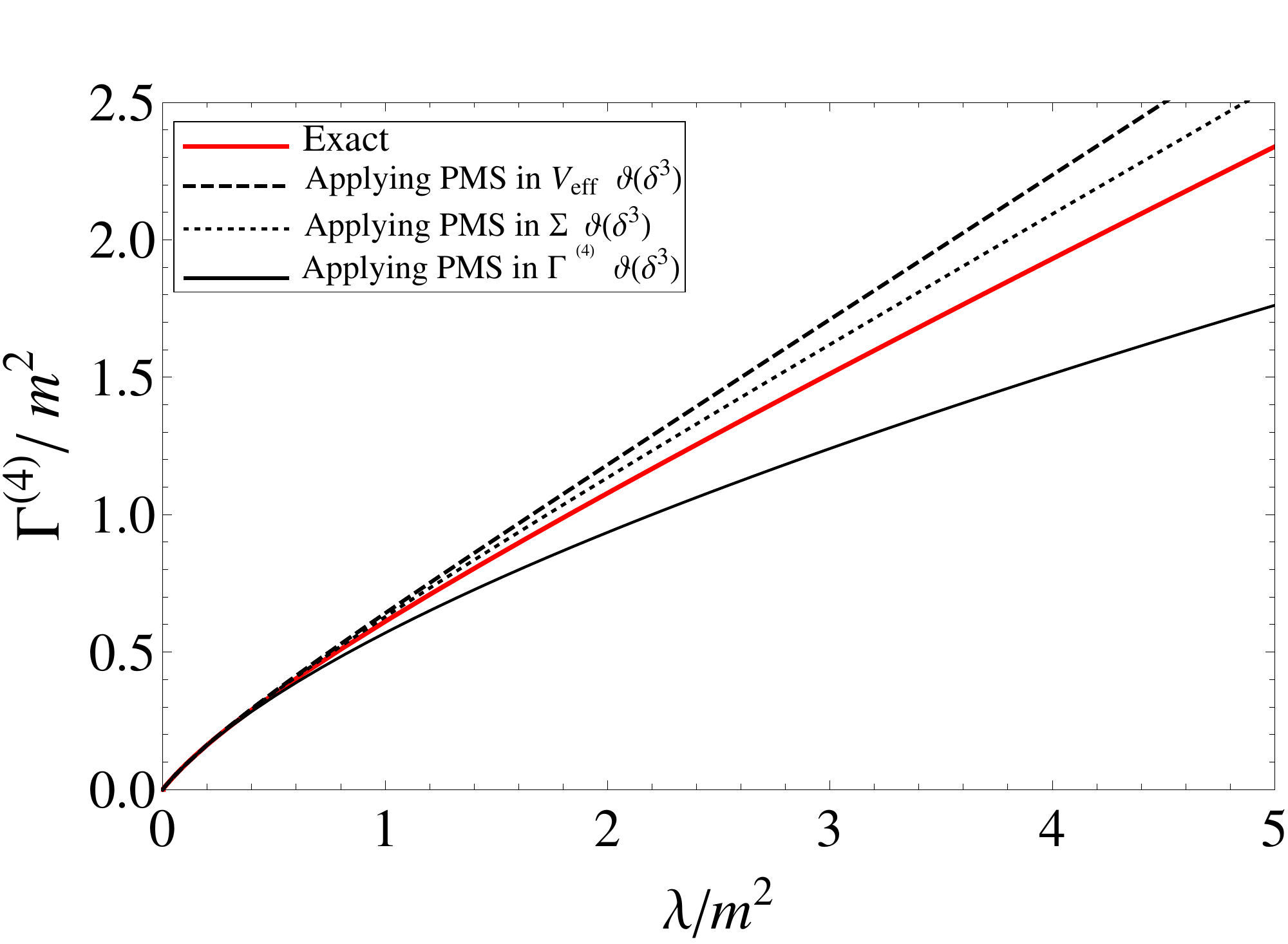}
\caption{Exact (solid line) and OPT results up to
   $\mathcal{O}(\delta^3)$ for the 1PI four point function
   $\Gamma^{(4)}$ at $N=2$. }
\label{fig10}
\end{center}
\end{figure}

In {}Fig.~\ref{fig8} we show how the different optimization schemes
within the OPT, the PMS, FAC and TP, affect the result for $V_{\rm
  eff}$. It is also compared the results by applying those
optimization schemes either to the effective potential itself, or to
the self-energy and using the produced optimal value of $\eta$ back in
the effective potential. These results show that the best agreement
with the exact solution is obtained by optimizing $\Sigma$ with PMS.
The same is repeated when we evaluate the self-energy $\Sigma$, whose
results are shown in {}Fig.~\ref{fig9}, and for the 1PI four point
function $\Gamma^{(4)}$,  shown in {}Fig.~\ref{fig10}.  In all these
cases, the best converging results are obtained when we choose to
optimize the self-energy in the PMS scheme.


\begin{center}
\begin{figure*}
\subfigure[]{\includegraphics[width=6.75cm,height=6cm]{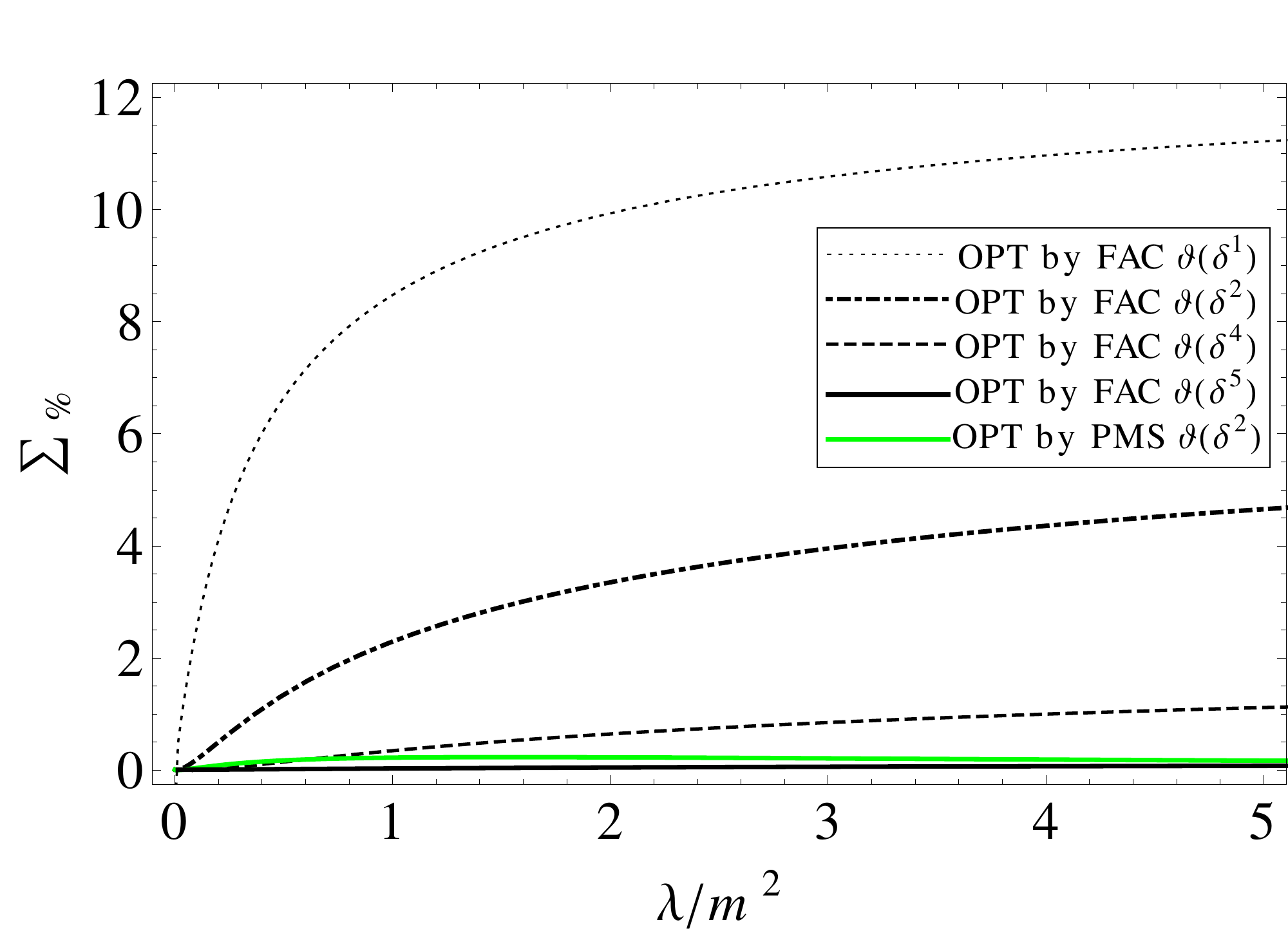}}\hspace{0.15cm}
\subfigure[]{\includegraphics[width=6.75cm,height=6cm]{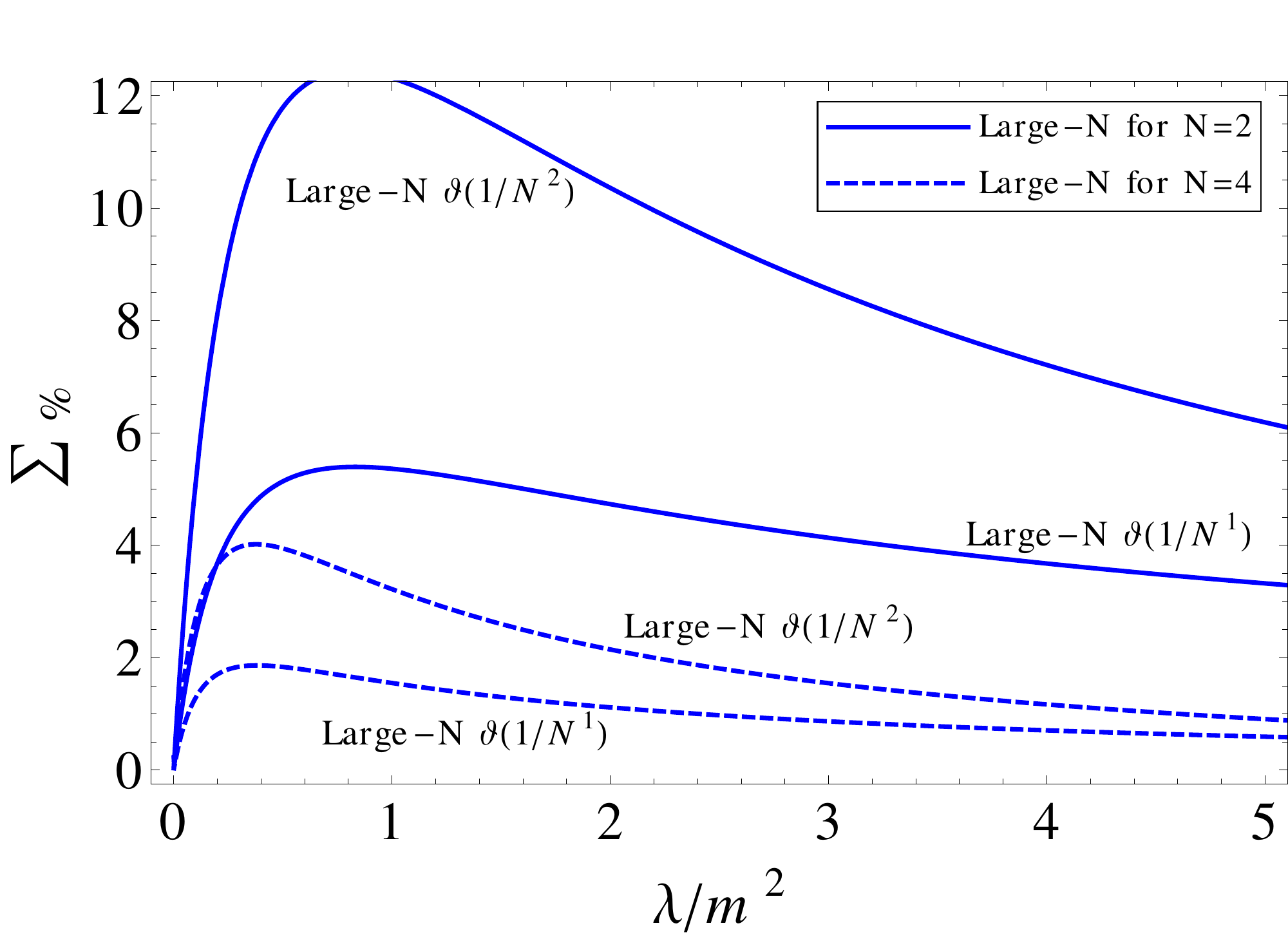}}
\caption{ Percentage difference for the $\Sigma$. Panel \textbf{(a)}:
   OPT results for $N=2$ obtained by optimizing $\Sigma$ by FAC  (up
   to $\mathcal{O}(\delta^5)$) and PMS (up to
   $\mathcal{O}(\delta^2)$).  Panel \textbf{(b)}: LN results for $N=2$
   and $N=4$.
}
\label{fig11}
\end{figure*}
\end{center}

\begin{center}
\begin{figure*}
\subfigure[]{\includegraphics[width=6.75cm,height=6cm]{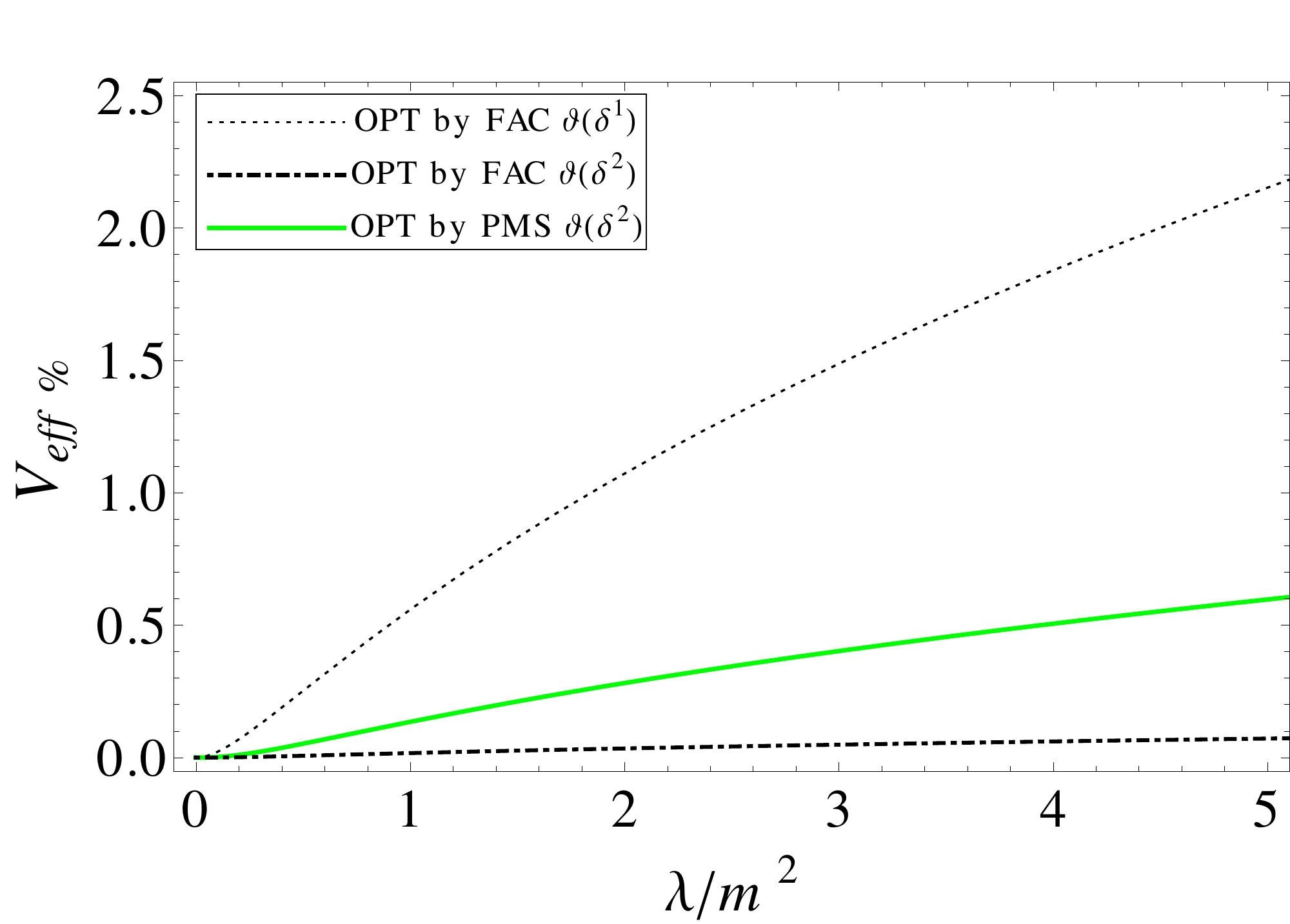}}\hspace{0.15cm}
\subfigure[]{\includegraphics[width=6.75cm,height=6cm]{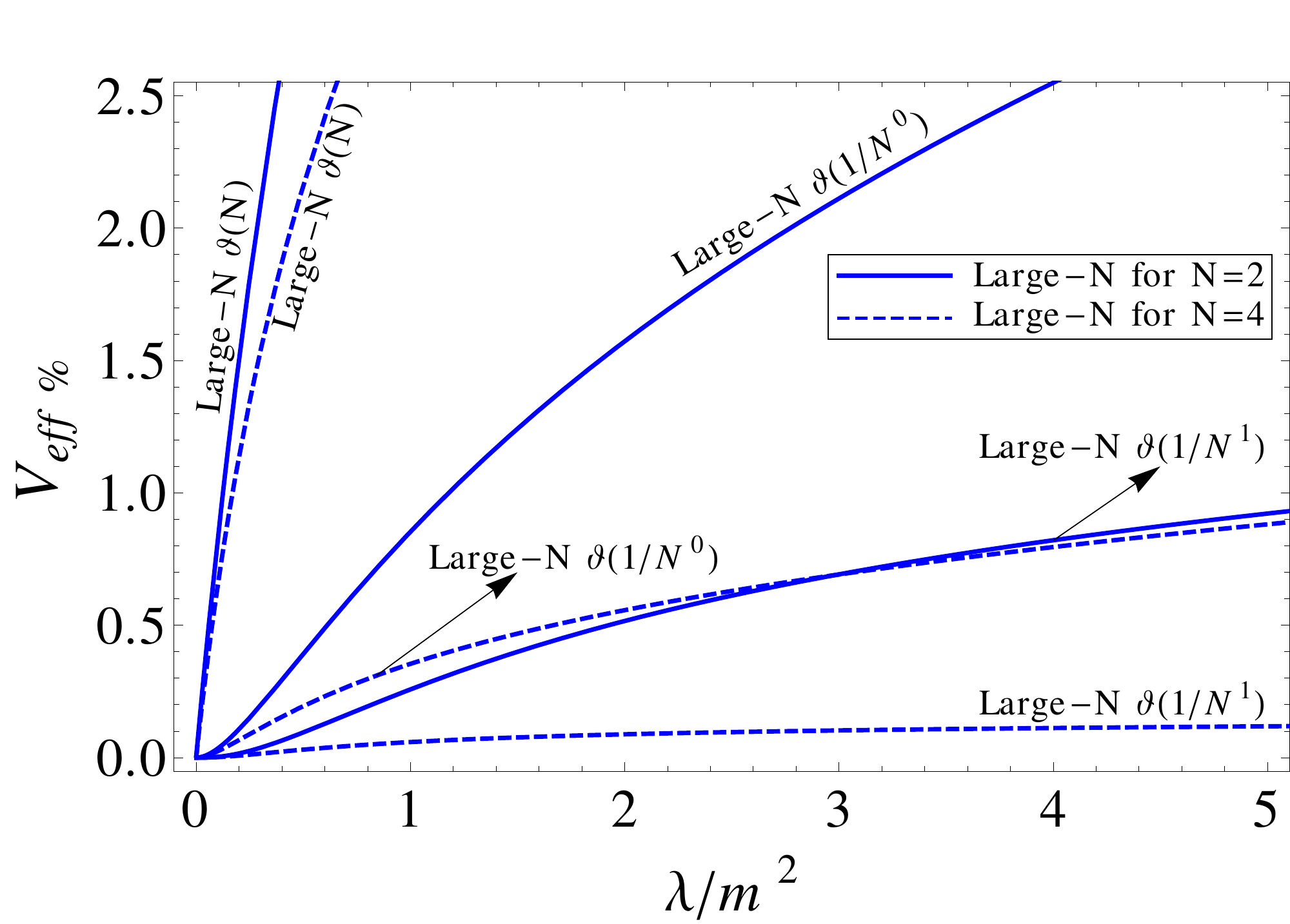}}
\caption{ Percentage difference for the $V_{\rm eff}$. Panel
   \textbf{(a)}: OPT results for $N=2$  obtained by optimizing
   $\Sigma$ by PMS and FAC. Panel \textbf{(b)}: LN results for $N=2$
   and $N=4$.
}
\label{fig12}
\end{figure*}
\end{center}

\begin{center}
\begin{figure*}
\subfigure[]{\includegraphics[width=6.75cm,height=6cm]{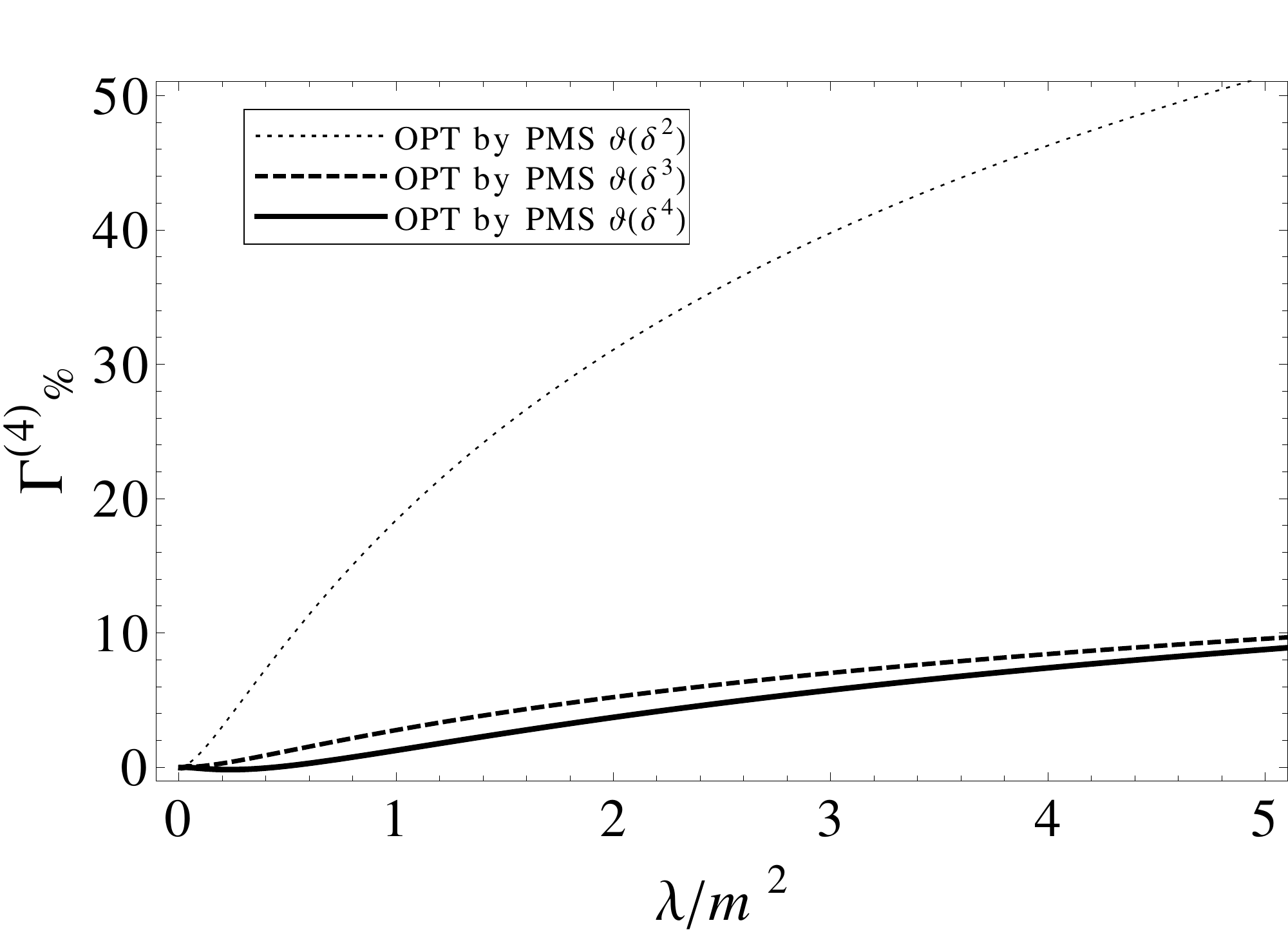}}\hspace{0.15cm}
\subfigure[]{\includegraphics[width=6.75cm,height=6cm]{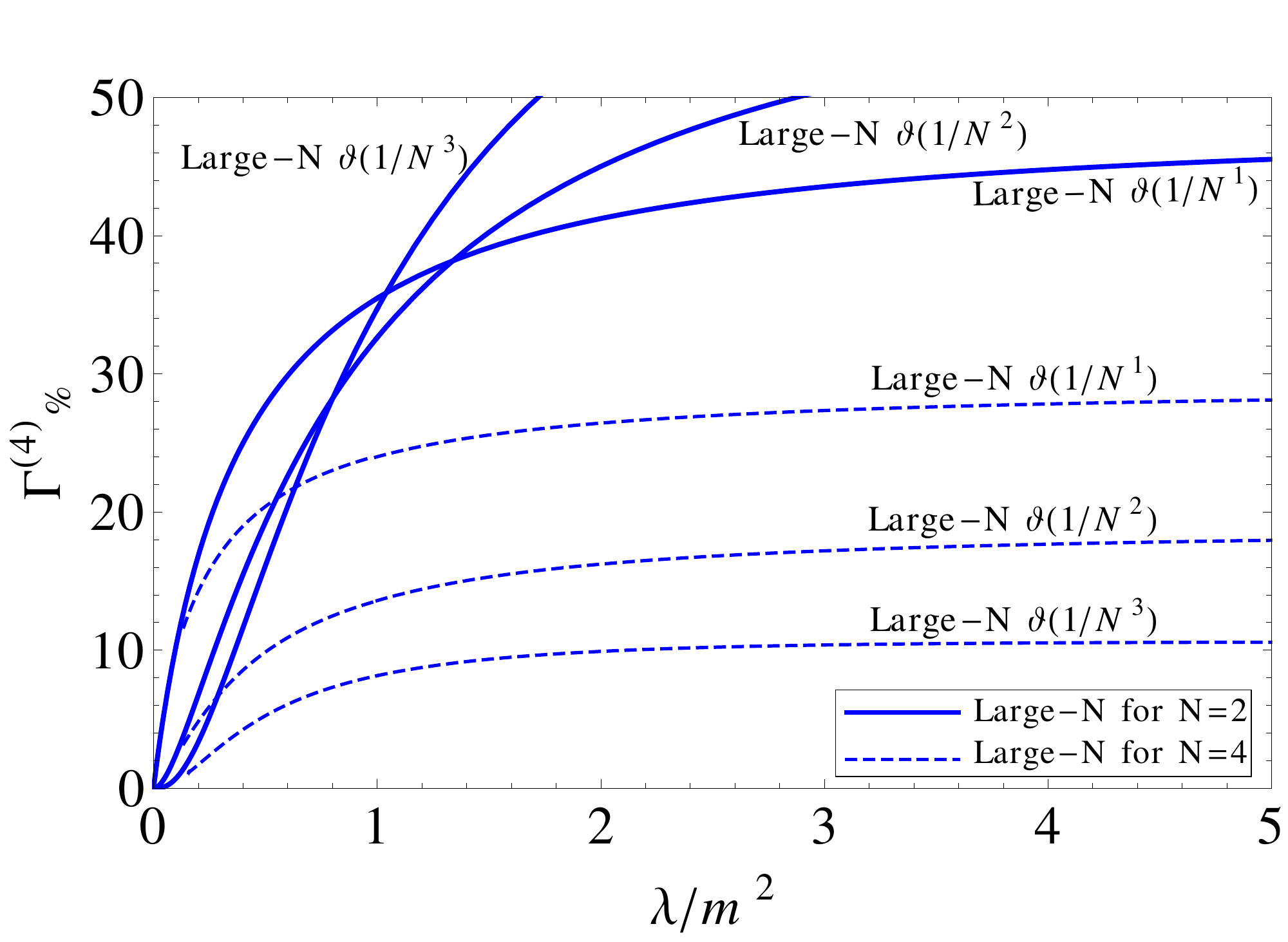}}
\caption{ Percentage difference for the $\Gamma^{(4)}$. Panel
   \textbf{(a)}: OPT results for $N=2$  obtained by optimizing
   $\Sigma$ by PMS. Panel \textbf{(b)}: LN results for $N=2$ and
   $N=4$.
}
\label{fig13}
\end{figure*}
\end{center}

Let us better quantify the differences between the two nonperturbative
methods studied in this work, the OPT and the LN
approximation. We want also to quantify the differences
between the different optimization
schemes. This is done next,
where we analyze the percentage difference that each method produces.
We define the percentage difference as

\begin{equation}
\Phi_\% = \left|\frac{\Phi_{\rm exact} -
  \Phi_{\rm approximated}}{\Phi_{\rm exact}}\right|,
\end{equation}
where $\Phi$ can represent any physical quantity evaluated in this
work: $V_{eff}$, $\Sigma$ or $\Gamma^{(4)}$.  In {}Figs.~\ref{fig11},
\ref{fig12} and \ref{fig13} we show the percentage difference for
$\Sigma$,  $V_{\rm eff}$ and $\Gamma^{(4)}$, respectively. Panel (a)
in these figures always refers to the OPT results, where, based on the
previous results, we have chosen to optimize the self-energy in the
two schemes that  performs well, the PMS and the FAC schemes.  The
results shown in panel (b) show the analogous percentage difference of
the approximation compared to the exact solution, but for the LN
approximation. {}For convenience, we have chosen two fixed values for
$N$, $N=2$ and $N=4$. We can see that the OPT  results are quite
impressive, showing good convergence in most cases already at second
order, while the  LN results present strong deviations from the exact
solution.  In particular, we can see that the OPT provides excellent
results for $\Sigma$ and $V_{\rm eff}$, but for $\Gamma^{(4)}$ it is
necessary to go to higher orders in $\delta$. As we expected, if we
increase the value of $N$ LN presents better results, but still it
under performs when compared with the OPT results.  


\section{Conclusions}
\label{conclusions}

In this work we have investigated in details the application of the
OPT in the 0-dimensional $O(N)$ scalar field model.  The questions we
wanted to answer with these work were, which optimization scheme works
better with the OPT and which quantity should we optimize to obtain
the optimum mass parameter that the OPT makes use to generate
nonperturbative results.  Through this study, we were able to better
access the convergence of the OPT with respect to the different
optimization schemes and with respect to each physical quantity that
they should be applied (for earlier studies on the convergence of the
OPT in the context of the anharmonic potential, see, e.g.,
Ref.~\cite{ldeqm}, while for a critical theory in field theory,
Ref.~\cite{Kneur:2002kq}).  Through the results obtained, we have
reached the conclusion that the PMS applied to the self-energy is in
general be best choice for producing results with better and faster
convergence.

One of the main advantages of the OPT method is its easy
implementation, that follows in practice the standard perturbative
expansion, but still able to generate nonperturbative results when
complemented with a proper optimization  procedure. In this work, we
have better clarified which optimization procedure is the ideal and to
which physical quantity it should be applied. {}For comparative
purposes, we have contrasted the results obtained with the OPT method
with another popular nonperturbative scheme, the LN expansion. Our
results have shown that the OPT is not only competitive but also out
performs the results obtained with the LN method.  Despite the
simplicity of the model we have used in this study, our results
indicate that the OPT method  can indeed be a better and simpler
alternative when applied to theories in physical dimensions $D>0$.

Our results are then expected to better motivate the use of the OPT
method as a powerful nonperturbative technique, specially when
combined with appropriate optimization schemes (for recent advances on
the OPT method and its combination also with renormalization group
techniques, see, e.g.,
Refs.~\cite{Kneur:2013coa,Kneur:2015moa,Kneur:2015uha}).
%
\section{Acknowledgements}
Work partially financed by Conselho Nacional de Desenvolvimento Cient\'{\i}fico e 
Tecnol\'ogico - CNPq, under the Grant Nos.  475110/2013-7 (R.L.S.F), 
232766/2014-2 (R.L.S.F), 308828/2013-5 (R.L.S.F) and 303377/2013-5 (R.O.R.), 
Funda\c{c}\~ao Carlos Chagas Filho de Amparo \`a Pesquisa do Estado do Rio de Janeiro (FAPERJ),
under grant No. E - 26 / 201.424/2014 (R.O.R.) and CAPES (D.S.R). R.L.S.F.  acknowledges 
the kind hospitality of the Center for Nuclear Research at Kent State where part of this work has been
done. 

\appendix
\section{Perturbation Theory}
\label{perturbationtheory}

We present here the results obtained when we use
perturbation theory to evaluate the Green's functions for the
0-dimensional $O(N)$ scalar field model. The usual strategy is to
expand the interaction term of the partition function  in powers of
the coupling constant $\lambda$ and then use hyperspherical
coordinates to evaluate the generating functional  at each order in
perturbation theory. {}For higher orders in the perturbative
expansion, we can  use the {}Feynman rules for the model (for futher
details, see, e.g., Ref.~\cite{Keitel:2011pn}).

The (non normalized) effective potential $V_{\rm eff}$ for this model
can be defined as  $V_{\rm eff} = - \ln Z$. The
effective potential evaluated in perturbation theory  up to
$\mathcal{O}\left(\frac{\lambda}{m^2}\right)^6$ is

\begin{eqnarray}
V_{\rm{eff}}& =&-\ln \left[\frac{2^{N-1} \pi^{N/2}}{m^{N/2}}\right] +
\frac{N^{2}+2N}{24}\left( \frac{\lambda}{m^{2}}\right)
-\frac{N^{3}+5N^{2}+6N}{144}\left( \frac{\lambda}{m^{2}}\right)
^{2}\nonumber\\ &+&\frac{5N^{4}+44N^{2}+128N^{2}+120N}{2592}\left(
\frac{\lambda}{m^{2}}\right) ^{3}\nonumber\\ &
-&\frac{7N^{5}+93N^{4}+468N^{3}+1040N^{2}+840N}{10368}\left(
\frac{\lambda}{m^{2}} \right) ^{4}\nonumber\\ &
+&\frac{21N^{6}+386N^{5}+2900N^{4}+11000N^{3}+20712N^{2}+15120N}{77760}
\left( \frac{\lambda}{m^{2}}\right) ^{5}\nonumber\\ & -&
\frac{33N^{7}+793N^{6}+8178N^{5}+45900N^{4}+146000N^{3}+245352N^{2}
+166320N}{279936}\nonumber\\
&\times& \left(
\frac{\lambda}{m^{2}}\right) ^{6} +\mathcal{O}\left(\left( \frac{\lambda}{m^{2}}\right)^{7}\right).
\end{eqnarray}

The results for the self-energy $\Sigma$ and $\Gamma^{(4)}$ are,
respectively, given by

\begin{eqnarray}
\frac{\Sigma}{m}&
=&\frac{N+2}{6}\left(\frac{\lambda}{m^{2}}\right)^{1}-\frac{N^{2}+6N+8}{36}
\left(
\frac{\lambda}{m^{2}}\right) ^{2} +
\frac{N^{3}+11N^{2}+38N+40}{108}\left( \frac{\lambda}{m^{2}}\right)
^{3}\nonumber\\  &
-&\frac{5N^{4}+84N^{3}+512N^{2}+1320N+1184}{1296}\left(
\frac{\lambda}{m^{2}}\right) ^{4}\nonumber\\ &
+&\frac{7N^{5}+163N^{4}+1492N^{3}+6640N^{2} + 14152N+11296}{3888}\left(
\frac{\lambda}{m^{2}}\right) ^{5}\nonumber\\   &
-&\frac{21N^{6}+638N^{5}+8020N^{4}+53000N^{3}+192292N^{2}+357680N+
261185}{23328}\nonumber\\
&\times&\left(
\frac{\lambda}{m^{2}}\right)^{6} + \mathcal{O}\left(\left( \frac{\lambda}{m^{2}}\right) ^{7}\right),
\end{eqnarray}
and

\begin{eqnarray}
\frac{\Gamma^{\left( 4\right)}}{m^{2}}&
=&\left(\frac{\lambda}{m^{2}}\right)^{1}-\frac{N+8}{6}
\left(\frac{\lambda}{m^{2}}\right)^{2}
+
\frac{3N^{2}+46N+140}{36}\left(\frac{\lambda}{m^{2}}\right)^{3}\nonumber\\ 
&-&\frac{5N^{3}+117N^{2}+772N+1536}{108}\left(
\frac{\lambda}{m^{2}}\right) ^{4}\nonumber\\ &
+&\frac{35N^{4}+1124N^{3}+11880N^{2}+51568N+79168}{1296}\left(
\frac{\lambda}{m^{2}}\right) ^{5}\nonumber\\ &
-&\frac{63N^{5}+2609N^{4}+38874N^{3}+271676N^{2}+906576N+1164032}{3888}
\left(\frac{\lambda}{m^{2}}\right) ^{6}\nonumber\\ &+&
\mathcal{O}\left(\left( \frac{\lambda}{m^{2}}\right) ^{7}\right).
\end{eqnarray}


\section{High order terms in the OPT}
\label{highorderopt}

In this appendix we present the results obtained for the OPT when
expanded up to order $\delta^{(5)}$ for effective potential $V_{ \rm
  eff}$, for the self-energy $\Sigma$ and for the 1PI four-point Green
function $\Gamma^{(4)}$. These results are  obtained following the
perturbative expansion in terms of the parameter $\delta$, using as
interaction term Eq.~(\ref{newinter}). The expressions for $V_{\rm
  eff}$, $\Sigma$ and $\Gamma^{(4)}$ are given, respectively, by
\begin{eqnarray}
V_{{\rm {eff}}} & = &
-\ln\left[\frac{2^{N-1}\ \pi^{N/2}}{(m+\eta)^{N/2}}\right]\delta^{0}+
\frac{\lambda(2N+N^{2})-
  12N\eta(m+\eta)}{24(m+\eta)^{2}}\delta^{1} \nonumber \\ & + &
\frac{-\lambda^{2}(6N+5N^{2}+N^{3})+\lambda\eta(24N+12N^{2})(m+\eta)-
\eta^{2}36N(m+\eta)^{2}}{144(m+\eta)^{4}}\delta^{2}
\nonumber \\ & + &
\left[\frac{\lambda^{3}(120N+128N^{2}+44N^{3}+5N^{4})-\lambda^{2}
\eta(432+36N^{2}+72N^{3})(m+\eta)}{2592(m+\eta)^{6}}\right.
  \nonumber \\ & + &
  \left.\frac{\lambda\eta^{2}(648N+324N^{2})(m+\eta)^{2}}{2592(m+\eta)^{6}}-
\frac{\eta^{3}N}{6(m+\eta)^{3}}\right]\delta^{3}\nonumber\\
&+&\left[\frac{-\lambda^{4}(840N+1040N^{2}+468N^{3}+93N^{4}+7N^{5})}
{10368(m+\eta)^{8}}\right.
  \nonumber \\ & + &
  \left.\frac{\lambda^{3}\eta(120N+128N^{2}+44N^{3}+5N^{4})}
{432(m+\eta)^{7}}-\frac{\lambda^{2}
    \eta^{2}(30N+25N^{2}+5N^{3})}{72(m+\eta)^{6}}\right.\nonumber\\
    &+&\left.
\frac{\lambda\eta^{3}(2N+N^{2})}{6(m+\eta)^{5}}-
  \frac{\eta^{4}N}{8(m+\eta)^{4}}\right]\delta^{4} \nonumber \\ 
& + &
\left[\frac{\lambda^{5}(15120N+20712N^{2}+11000N^{3}+2900N^{4}+386N^{5}+
21N^{6})}{77760(m+\eta)^{10}}\right.
  \nonumber \\ & - &
  \left.\frac{\lambda^{4}\eta(840N+1040N^{2}+468N^{3}+93N^{4}+7N^{5})}
{1296(m+\eta)^{9}}\right.\nonumber\\
& -&\left. 
  \frac{\lambda^{2}\eta^{3}(30N+25N^{2}+5N^{3})}{36(m+\eta)^{7}} +  \frac{\lambda^{3}\eta^{2}(840N+896N^{2}+308N^{3}+35N^{4})}
{864(m+\eta)^{8}}\right.\nonumber\\
&+&\left.
  \frac{\lambda\eta^{4}(10N+5N^{2})}{24(m+\eta)^{6}}-
\frac{\eta^{5}N}{10(m+\eta)^{5}}\right]\delta^{5} +  \mathcal{O}\left(\delta^{(6)}\right),
\end{eqnarray}

\begin{eqnarray}
\frac{\Sigma}{m} & = &
\frac{\eta}{m}\delta^{0}+\frac{\lambda(2+N)-6\eta(m+\eta)}{6m(m+\eta)}
\delta^{1}\nonumber\\
&-&\frac{\lambda^{2}(8+6N+N^{2})-\lambda\eta(12N+6N)
(m+\eta)^{2}}{36m(m+\eta)^{3}}\delta^{2}
\nonumber \\ 
& + & \frac{\lambda^{3}(40+38N+11N^{2}+N^{3})}{108m(m+\eta)^{5}}\delta^{3}
\nonumber \\ 
& - & \frac{\lambda^{2}\eta(72+54N+9N^{2})(m+\eta)+
  \lambda\eta^{2}(36+18N)(m+\eta)^{2}}{108m(m+\eta)^{5}}\delta^{3}
\nonumber \\ 
& - &\left[\frac{\lambda^{4}(1184+1320N+512N^{2}+84N^{3}+5N^{4})}{1296m(m+\eta)^{7}}\right.  \nonumber \\ 
& + &\left.\frac{ \lambda^{3}
\eta(2400+2280N+660N^{2} + 60N^{3})(m+\eta)}{1296m(m+\eta)^{7}}\right.  \nonumber \\ 
& + & \left.\frac{\lambda^{2}\eta^{2}(8+6N+N^{2})}{6m(m+\eta)^{5}}-
  \frac{\lambda\eta^{3}(2+N)}{6m(m+\eta)^{4}}\right]\delta^{4}\nonumber\\
 & - & \left[\frac{\lambda^{5}(11296+14152N+6640N^{2}+1492N^{3}+163N^{4}+7N^{5})}{3888m(m+\eta)^{9}}\right.
  \nonumber \\ & - &
  \left. 
\frac{\lambda^{4}\eta(8288+9240N+3584N^{2}+588N^{3}+
    35N^{4})}{1296m(m+\eta)^{8}}\right.  \nonumber \\ & + &
  \left.\frac{\lambda^{3}\eta^{2}(200+190N+55N^{2}+5N^{3})}{36m(m+\eta)^{7}}-
  \frac{\lambda^{2}\eta^{3}(40+30N+5N^{2})}{18m(m+\eta)^{6}}\right.\nonumber\\
  &+&
\left.\frac{\lambda\eta^{4}(2+N)}{6m(m+\eta)^{5}}\right]\delta^{5}+
\mathcal{O}\left(\delta^{(6)}\right),
\end{eqnarray}

and

\begin{eqnarray}
\frac{\Gamma^{\left(4\right)}}{m^{2}} & = &
\frac{\lambda}{m^{2}}\delta^{1}-\frac{\lambda^{2}(8+N)}{6m^{2}(m+\eta)^{2}}
\delta^{2} \nonumber \\
&+& \frac{\lambda^{3}(140+46N+3N^{2})-
\lambda^2\eta(96+12N)(m+\eta)}{36m^{2}(m+\eta)^{4}}\delta^{3}
\nonumber \\ 
& - &
\left[\frac{\lambda^{4}(1536+1772N+117N^{2}+5N^{3})}{108m^{2}(m+\eta)^{6}}-
\frac{\lambda^{3}\eta(140+46N+3N^{2})}{9m^{2}(m+\eta)^{5}} \right.\nonumber\\
&+& \left. \frac{\lambda^{2}
\eta^{2}(8+N)}{2m^{2}(m+\eta)^{4}}\right]\delta^{4}
\nonumber \\ 
&+ &
\left[\frac{\lambda^{5}(79168+51568N+11880N^{2}+1124N^{3}+35N^{4})}
{1296m^{2}(m+\eta)^{8}}\right.\nonumber\\
& - &\left. \frac{\lambda^{4}\eta(1136+772N+1117N^{2}+5N^{3})}
{18m^{2}(m+\eta)^{7}}\right.
\nonumber \\ &
  + &
  \left.\frac{\lambda^{3}\eta^{2}(700+230N+15N^{2})}{18m^{2}(m+\eta)^{6}}-
\frac{\lambda^{2}\eta^{3}(16+2N)}{3m^{2}(m+\eta)^{5}}\right]\delta^{5}\nonumber\\
& + & \mathcal{O}\left(\delta^{(6)}\right).
\end{eqnarray}


\section{Large-$N$ Approximation}
\label{sectionlargen}

The LN approximation applied for the 0-dimensional $O(N)$ scalar
field model was described in details in Ref.~\cite{Keitel:2011pn}. In
this appendix we reproduce some of the expressions obtained from that
reference that we have used in this work. {}For this model, we have
that $\boldsymbol{\varphi}^2 = \mathcal{O}(N)$ and
$\lambda=\mathcal{O}(1/N)$, which shows that $\lambda \rightarrow
\tilde{\lambda}/N$ is a reasonable replacement. In the large-$N$ limit
we can evaluate  the partition function $Z$ in the saddle-point
approximation~\cite{Arfken} (leading and next-to-leading orders
terms). 

Performing the change of variables $\left(y =
\frac{\boldsymbol{\varphi}^{2}}{N}\right)$ and using hyperspherical
coordinates we can evaluate the partition function that can be written
as 

\begin{equation}
Z =  \Omega_{N}N^{N/2}\int_{0}^{\infty} \frac{dy}{2y} e^{-N f(y)},
\end{equation}
where the function $f(y)$ is defined by 

\begin{equation}
f(y)  =
\frac{m}{2}y+\frac{\tilde{\lambda}}{4!}y^{2}-\frac{1}{2}\ln(y). 
\label{fylargen}
\end{equation}
In a saddle point approximation we perform an expansion in
Eq.~(\ref{fylargen}) around its minimum,

\begin{equation}
y_{0}  =  \frac{3m}{\tilde{\lambda}}\left(
\sqrt{1+\frac{2\tilde{\lambda}}{3m^{2}}}-1\right).
\end{equation}
Expanding around this minimum and performing the integral, we obtain
the partition function~\cite{Keitel:2011pn}

\begin{eqnarray}
Z & = & \Omega _{N}N^{\frac{N}{2}}e^{ -N f\left( y_{0}\right)}\left(
\frac{1}{4y_{0}^{2}}\frac{2\pi }{f^{\prime \prime } \left(
  y_{0}\right) }\right) ^{1/2}  \nonumber
\\       &\times&\left[1+\frac{12m^{2}y_{0}^{2}-27my_{0}+16}
{6N(2-my_{0})^{3}}\right]\left[1+\mathcal{O}(1/N)\right], 
\end{eqnarray}
with

\begin{equation}
f(y_{0}) = \frac{m y_{0}}{4}+\frac{1}{4}-\frac{1}{2}\ln(y_{0}).
\end{equation}

Next-to-leading order LN results for $\Gamma^{(0)}$ can be obtained by
taking the logarithm of $\frac{Z}{Z_0}$:

\begin{eqnarray}
\Gamma^{(0)} & = & \left[\frac{m y_0}{4} - \frac{1}{4} -
  \frac{1}{2}\ln(m y_0)\right] N^1 + \frac{1}{2}\ln(2 - m y_0) N^0
\nonumber\\ &-& \frac{(8 + m y_{0})(m y_0 - 1)^{2}}{6(2-m
  y_0)^3}\frac{1}{N^{1}} \nonumber\\ &+&\mathcal{O}(1/N^{2}).
\end{eqnarray}

Higher order terms in $1/N$ usually are very difficult to obtain
because we need go beyond the saddle-point approximation,  including
fluctuations in the corrections. But for the case of the 0-dimensional
$O(N)$ scalar field model,  it can be obtained by successive
derivatives of $\Gamma^{(0)}$ with respect to $m$.  {}Following this
procedure, we can for the self-energy the result

\begin{eqnarray}
\Sigma & = & \left[\frac{1}{y_{0}}-m\right]\frac{1}{N^{0}}  +
\left[\frac{2(1-my_{0})}{y_{0}(2-my_{0})^{2}}\right]
\frac{1}{N^{1}}\nonumber\\ &+&
\left[\frac{4(my_{0}-1)^{2}(3my_{0}-1)}{y_{0}(2-my_{0})^{5}}\right]
\frac{1}{N^{2}}
\nonumber\\ &+&\mathcal{O}(1/N^{3}), 
\end{eqnarray}
and also for the 1PI four-point function,

\begin{eqnarray}
\Gamma^{(4)} &=&
\left[\frac{6(1-my_{0})}{y_{0}^{2}(2-my_{0})}\right]\frac{1}{N}
\nonumber\\ &-&
\left[\frac{12(1-my_{0})^{2}(m^{2}y_{0}^{2}-3my_{0}+6)}{y_{0}^{2}(2-my_{0})^4}
\right]
\frac{1}{N^{2}}  \nonumber\\ &+ &\left[\frac{24(1-my_{0})^{3}}
  {y_{0}^{2}(2-my_{0})^{7}}\right.\nonumber\\
  &\times& \left.(m^{4}y_{0}^{4}-8m^{3}y_{0}^{3}+35m^{2}y_{0}^{2}- 49my_{0}+56)\right] \frac{1}{N^{3}} \nonumber\\ &+&
\mathcal{O}(1/N^{4}).
\end{eqnarray}

\section{The optimum $\eta$}
\label{appD}

We give below the expressions for the optimum $\eta$ obtained
in each of the optimization procedures that we have explained in the text,
when applied to the different physical quantities, i.e., to the self-energy,
the effective potential and to the four-point Green's function.

\begin{enumerate}
\item $\bar{\eta}$ obtained by optimizing the self-energy $\Sigma$:

\begin{itemize}

\item using PMS:

\begin{eqnarray}
 \bar{\eta} &=& \frac{m}{2} \left(-1+\sqrt{1+(4+N)\frac{\lambda }{m^2}}\right)\delta^{2} 
 +  \frac{m}{6} \biggl(-3 \nonumber\\
 &+&\sqrt{9 +12(4+N) \frac{\lambda }{m^2}  - 2\sqrt{6\left(-4+3 N+N^2\right) 
 \frac{\lambda ^2}{m^4}}}\biggr)\nonumber\\ &\times& \delta^{3} + \mathcal{O}\left(\delta^{4}\right).
\end{eqnarray}
Note that there is no solution (real and positive) for $\bar{\eta}$ at order $\delta^1$.

\item using FAC:

 \begin{eqnarray}
 \bar{\eta} &=& \frac{m}{6} \left(-3  + \sqrt{9 +6(2+N) \frac{\lambda }{m^2}}\right)\delta^{1} 
 \nonumber\\
 &+&
 \frac{m}{6} \left(-3  + \sqrt{9 +6(4 +N)\frac{\lambda }{m^2}}\right) \delta^{2}\nonumber \\
 &+&    \frac{m}{6} \left(-3 
 +\sqrt{9 +9(4+N) \frac{\lambda }{m^2} - 3\sqrt{\left(N^2-16\right) \frac{\lambda ^2}{m^4}}}\right)
 \delta^{3}\nonumber \\ 
 &+& \mathcal{O}\left(\delta^{4}\right).
\end{eqnarray}
Note that for $N<4$ there is no solution (real and positive) for $\bar{\eta}$ at order $\delta^3$.

\item using TP:

\begin{eqnarray}
 \bar{\eta}  &=& \frac{m}{4} \left(-1+\sqrt{9 +8(4+N) \frac{\lambda }{m^2}}\right)\delta^{2} 
 + \frac{1}{12}  \Biggl\{\sqrt{F} - 3 m  \nonumber\\
 &-& \left[\frac{432 B-\left(\sqrt{F}-3 m\right) \left(-24 A+F+3 \sqrt{F} m-18 m^2\right)}{\sqrt{F}}
 \right]^\frac{1}{2}\Biggr\}  \nonumber \\
 &\times& \delta^{3} + \mathcal{O}\left(\delta^{4}\right),
\end{eqnarray}
where we have defined 
\begin{eqnarray}
 A&=&3 m^2 + 4\lambda(4  +  N ),  \\
 B&=& m^3+m \lambda(4 +N ),  \\
 C&=&  (4 + N) \left[6 m^2 \lambda + 5 \lambda^2 (5+N)\right],  \\
 D &=& 36 m^4 + (4+ N )\left[87m^2 \lambda +2 \lambda^2 (107+23 N)\right],  \\
 E &=&  -4 A^3+1458 B^2+72 A C-162 A B m+81 C m^2,\nonumber \\
 &+&\biggl\{-16 D^3+\biggl[-4 A^3+18 A (4 C-9 B m)+81 \biggl(18 B^2 \nonumber\\
 &+& C m^2\biggr)\biggr]^2\biggr\}^{\frac{1}{2}},  \\
 F&=& 8 A+\frac{2^{8/3} D}{E^{1/3}}+2^{4/3} E^{1/3}+9 m^2. 
\end{eqnarray}
Note that in this case there is no solution (real and positive) for $\bar{\eta}$ at order $\delta^1$.

\end{itemize}

\item $\bar{\eta}$ optimizing the effective potential $V_{\rm{eff}}$

\begin{itemize}

\item using PMS:

\begin{eqnarray}
\bar{\eta} &=& \frac{m}{6} \left(-3 + \sqrt{9 +(12+6N)\frac{\lambda }{m^2}}\right)\delta^{1} +
\frac{m}{6} \Biggl(\frac{}{}-3 \nonumber\\ 
&+& \sqrt{9 +(18+9N) \frac{\lambda ^2}{m^4}-3 \sqrt{\left(-12-4 N+N^2\right) \frac{\lambda ^2}{m^4}}}
\Biggr)\delta^{2} \nonumber \\
&+& \frac{1}{6 G^{1/3}}\left(-3 G^{1/3} m + G^\prime\right)\delta^{3}+ \mathcal{O}\left(\delta^{4}\right),
\end{eqnarray}
\noindent where we have defined the constants
\begin{eqnarray}
G&=&9\left(8+N^3\right) \lambda ^3 + \biggl[-3(2+N)^2 \biggl(-2160+1296 N-36 N^2,
\nonumber\\
 &-& 116 N^3+5 N^4\biggr) \lambda ^6\biggr]^{\frac{1}{2}},
\end{eqnarray}
and 
\begin{eqnarray}
G^\prime &=& \biggl[ 2^{2/3} 3^{1/3} G + 2^{7/3} 3^{2/3} G^{1/3}\left(-6-N+N^2\right) \lambda ^2, 
\nonumber\\
&+& G^{2/3} \left(9 m^2+ (24+12N) \lambda \right)\biggr]^{\frac{1}{2}}.
\end{eqnarray}
Note that in this case, for $N<6$ there is no solution (real and positive) for $\bar{\eta}$ at order $\delta^2$.

\item using FAC:

\begin{eqnarray}
 \bar{\eta} &=& \frac{m}{6} \left(-3 + \sqrt{9 +3(2+N) \frac{\lambda }{m^2}}\right)\delta^{1} \nonumber\\ 
 &+& \frac{1}{6 H^{1/3}}\left(-3 H^{1/3} m+ H^\prime \right)\delta^{3} 
 + \mathcal{O}\left(\delta^{4}\right),
\end{eqnarray}
\noindent where we have defined the additional constants
\begin{eqnarray}
H&=&\left(24+4 N-2 N^2+N^3\right) \lambda ^3, \nonumber\\
&+& \sqrt{8(2+N)^2 \left(72-12 N-4 N^2+N^3\right) \lambda ^6},
\end{eqnarray}
and
\begin{eqnarray}
H^\prime &=& \biggl[ 3I+H^{1/3}\left(-12-4 N+N^2\right)3 \lambda ^2 \nonumber\\
&+& 9 H^{2/3}m^2 \left(1+(2+N) \frac{\lambda}{m^2} \right)\biggr]^{\frac{1}{2}}.
\end{eqnarray}
Here there is no solution (real and positive) for $\bar{\eta}$ at order $\delta^2$.

\item using TP:

\begin{eqnarray}
\bar{\eta} &=& \frac{m}{\sqrt{2}}\left(\sqrt{2 +(2 +N)\frac{\lambda }{m^2}}\right) \delta^{1} \nonumber \\
&+&\left( -\frac{1}{2} \sqrt{2 I+\frac{6 \sqrt{6} J}{\sqrt{54 I+\frac{27 L}{O^{1/3}}+O^{1/3}}}-\frac{L}{2 O^{1/3}}-
\frac{O^{1/3}}{54}}\right.\nonumber\\&+&\left.\frac{\sqrt{54 I+\frac{27 L}{O^{1/3}}+O^{1/3}}}{6 \sqrt{6}}\right)\delta^{2} + \mathcal{O}\left(\delta^{3}\right),
\end{eqnarray}
\noindent where
\begin{eqnarray}
 I&=& 2 m^2+ \lambda ( 2  +N ),  \\
 J&=&2 m^3+m \lambda(2 +N) ,  \\
 K&=& -( 2 + N) \biggl[ 9m^2 \lambda +5\lambda ^2  (3+N)\biggr],  \\
 L &=& 108 m^4+  (2+ N)  \biggl[ 180 m^2 \lambda \nonumber\\ 
 &+&\lambda ^2  (174 + 67 N) \biggr], \\
 M &=& 2916 m^4+ 27 (2+N) \biggl[180 m^2 \lambda \nonumber\\
 &+&\lambda ^2 (174+67 N) \biggr], \\
 O&=& -19683 I^3+78732 J^2-17496 I K \nonumber\\
 &-& \sqrt{4782969 \left(9 I^3-36 J^2+8 I K\right)^2- M^3}.
\end{eqnarray}

\end{itemize}

\item $\bar{\eta}$ optimizing the $\Gamma^{(4)}$

\begin{itemize}
\item using PMS:

  \begin{eqnarray}
\bar{\eta} &=&  \frac{m}{18 (8+N)}\biggl\{-9(8+N)+
\biggl[ 9(8+N)  \biggl(9 (8+N) \nonumber \\ 
&+&  4  \left(140+46 N+ N^2\right)\frac{\lambda }{m^2}\biggr)\biggr]^{\frac{1}{2}}\biggr\}\delta^{3} + \mathcal{O}\left(\delta^{4}\right).
\end{eqnarray}

\item using FAC:

\begin{eqnarray}
\bar{\eta} &=& \frac{m}{24 (8+N)}\biggl\{-12(8+N)+
\biggl[48(8+N)\biggl(3 (8 + N)\nonumber\\ 
&+& \left(140 +46 N +3 N^2 \right)\frac{\lambda }{m^2}\biggr)\biggr]^{\frac{1}{2}}\biggr\} \delta^{3} + \mathcal{O}\left(\delta^{4}\right).
\end{eqnarray}

\item using TP:

 \begin{eqnarray}
\bar{\eta} &=& \frac{m}{9 (8+N)}\biggl\{-3(8+N)+\biggl[3(8+N)\biggl(12(8+N) \nonumber\\
&+&  5  (140+46N+3 N^{2})\frac{\lambda }{m^2}\biggr)\biggr]^{\frac{1}{2}}\biggr\}\delta^{3}
+ \mathcal{O}\left(\delta^{4}\right),
\end{eqnarray}

\end{itemize}
\noindent Note that there is no solution (real and positive) for $\bar{\eta}$ at orders $\delta^1$ and $\delta^2$ 
for the three optimization procedures applied to $\Gamma^{(4)}$.

\end{enumerate}


\end{document}